\documentclass[apj]{emulateapj}
\usepackage{graphicx}

\newcommand{\HI}{H$\;${\small\rm I}\relax}
\newcommand{\HII}{H$\;${\small\rm II}\relax}
\newcommand{\htwo}{H$_2$}
\newcommand{\Ha}{H$\alpha$\relax}
\newcommand{\halpha}{H$\alpha$\relax}
\newcommand{\Nh}{\ensuremath{N_H}\relax}
\newcommand{\nH}{\ensuremath{n_{\rm H}}}
\newcommand{\z}{$z$}
\newcommand{\sdssr}{$r$}
\newcommand{\msun}{M$_\odot$}

\newcommand{\etal}{{et al.}}

\newcommand{\percc}{cm$^{-3}$}

\newcommand{\e}[1]{10^{#1}}

\newcommand{\garciaburillo}{Garc\'{\i}a-Burillo}

\newcommand{\wfpc}{WFPC2}
\newcommand{\hst}{{\em HST}}

\newcommand{\pI}{HS97}
\newcommand{\pII}{HS99}
\newcommand{\pIII}{HS00}
\newcommand{\ths}{THS04}
\newcommand{\ngc}{NGC~}

\slugcomment{Accepted for publication in AJ}
\shortauthors{Rueff et al. 2012}

\begin{document}

\title{The Relationship Between the Dense Neutral and Diffuse
Ionized Gas in the Thick Disks of Two Edge-On Spiral Galaxies \altaffilmark{1}}

\author{Katherine M. Rueff\altaffilmark{2}, J. Christopher Howk\altaffilmark{2},
Marissa Pitterle\altaffilmark{2}, Alec S. Hirschauer\altaffilmark{2,3}, 
Andrew J. Fox\altaffilmark{4}, \and Blair D. Savage\altaffilmark{5}}

\altaffiltext{1}{Based on Observations obtained with the NASA/ESA
  Hubble Space Telescope operated at the Space Telescope Science
  Institute, which is operated by the Association of Universities for
  Research in Astronomy, Inc., under NASA contract NAS5-26555.
  Also, based on data acquired using the Large Binocular Telescope (LBT).
  The LBT is an international collaboration among institutions in the US, 
  Italy, and Germany. LBT Corporation partners are the University of Arizona, 
  on behalf of the Arizona university system; Instituto Nazionale do Astrofisica, 
  Italy; LBT Beteiligungsgesellschaft, Germany, representing the Max Planck Society, 
  the Astrophysical Institute of Postdam, and Heidelberg University; Ohio State
  University, and the Research Corporation, on behalf of the University of Notre Dame, the
  University of Minnesota, and the University of Virginia.  Also, based on observations
  obtained by the WIYN Observatory which is a joint facility of the University of
  Wisconsin-Madison, Yale University, Indiana University, and the National Optical
  Astronomy Observatories.}
\altaffiltext{2}{Department of Physics, University of Notre Dame, 
Notre Dame, IN 46556. krueff@nd.edu}
\altaffiltext{3}{Department of Astronomy, Indiana University, Bloomington, IN 47405.}
\altaffiltext{4}{Space Telescope Science Institute, 3700 San Martin Drive, 
Baltimore MD 21218.}
\altaffiltext{5}{Department of Astronomy, University of Wisconsin,
  Madison, Madison, WI 53706.}
  
\begin{abstract}
We present high-resolution, optical images (BVI $+$ \Ha) of the multiphase interstellar
medium (ISM) in the thick disks of the edge-on spiral galaxies \ngc 4013 and \ngc 4302.
Our images from the {\em Hubble Space Telescope} (\hst), Large Binocular Telescope, and
WIYN 3.5-m reveal an extensive population of filamentary dust absorption seen to \z\ $\sim
2-2.5$ kpc. Many of these dusty thick disk structures have characteristics reminiscent of
molecular clouds found in the Milky Way disk. Our \Ha\ images show the extraplanar diffuse
ionized gas (DIG) in these galaxies is dominated by a smooth, diffuse component.  The
strongly-filamentary morphologies of the dust absorption have no counterpart in the
smoothly distributed \Ha\ emission. We argue the thick disk DIG and dust-bearing filaments
trace physically distinct phases of the thick disk ISM, the latter tracing a dense, warm
or cold neutral medium. The dense, dusty matter in the thick disks of spiral galaxies is
largely tracing matter ejected from the thin disk via energetic feedback from massive
stars. The high densities of the gas may be a result of converging gas flows.  This dense
material fuels some thick disk star formation, as evidenced by the presence of thick disk
\HII\ regions. 
\end{abstract}

\section{Introduction}
The feedback of energy and matter from stars to the interstellar medium (ISM) plays a
central role in the evolution of galaxies.  It impacts the physical/thermal state of the
ISM, affecting the ability of the gas to form stars and affects the galactic-scale star
formation by ejecting material from the thin disk or altogether from a galaxy (e.g., Dekel
\& Silk 1986). Most disk galaxies with sufficient star formation rates (or star formation
rate surface densities), show extended ``thick disks'' of warm interstellar matter
extending $\sim 2$ kpc away from the galaxy mid-plane (Rossa \& Dettmar 2003a, 2003b). The
presence of a thick disk ISM supported against gravity is a direct result of energetic
feedback processes (although some additional material may be contributed through infalling
matter; see Sancisi et al. 2008).  Thus, the study of the statistical properties (e.g.,
Rossa \& Dettmar 2003a, 2003b; Howk \& Savage 1999) and detailed physics (e.g., Haffner et
al. 2009) of the thick disk ISM can help us understand the role that feedback plays in
disk galaxies.

The energetic processes that expel gas from the thin disk also act on the solid phase of
the ISM, interstellar dust grains. This has been utilized to study the physical structure
as well as the phase structure of the thick disk ISM (Howk \& Savage 1997, 1999, 2000;
Alton \etal\ 2000, Thompson \etal\ 2004).  While the energetic expulsion of matter from
the disk can impact the dust-to-gas ratio through dust destruction, the fact that
prevalent extraplanar dust is seen in $\sim50\%$ of normal spiral galaxies (Rossa \&
Dettmar 2003a; Howk \& Savage 1999) and even in galactic superwinds such as seen in M82
(Kaneda et al. 2010; Roussel et al. 2010; Engelbracht et al. 2006; Hoopes et al. 2005;
Scarrott et al. 1991) demonstrates that dust can survive the ejection from the thin disk. 
In fact, the detection of thick disk dust via the extinction it produces against
background starlight in direct broadband images of edge-on galaxies is the most
observationally efficient method for detecting extraplanar interstellar matter (Howk \&
Savage 2000).

The first detailed discussion of the implications of extraplanar dust was in the Howk \&
Savage (1997) (hereafter HS97) study of the nearby edge-on spiral galaxy \ngc 891,
although the presence of high-\z\ dust clouds in this galaxy were noted earlier (Keppel
\etal\ 1991; Sandage 1961).  \pI\ showed that the thick disk of \ngc 891 was threaded with
a network of dusty clouds and filaments observed to heights up to $z \sim 2$ kpc.  The
most remarkable implication of this work was that the individual dust-bearing clouds have
very large masses, often in excess of $10^5 M_{\odot}$.  Howk \& Savage (1999) (hereafter
HS99) subsequently surveyed a group of nine nearby edge-on spiral galaxies, demonstrating
that high-\z\ dust is found in many galaxies and is correlated with the presence of
extraplanar diffuse ionized gas (DIG) seen in \Ha\ imaging.  Subsequently, Rossa \&
Dettmar (2003a, 2003b) showed that $\sim40\%$ of a much larger sample of spiral galaxies
exhibit extraplanar dust and verified the correlation between the presence of extraplanar
dust and DIG emission.  \pII\ and Howk \& Savage (2000; hereafter \pIII) argued that the
gas traced by extraplanar extinction features represented a much denser medium than that
seen in \Ha\ emission, with densities an order of magnitude or more higher than DIG
material (see also Keppel \etal\ 1991). The extraplanar dust clouds and filaments, which
are not spatially correlated with the DIG, are only visible because they have
significantly higher extinction -- and likely higher column densities and particle
densities -- than their surroundings.  \pIII\ hypothesized that the dust structures trace
a cold, neutral medium (CNM) in the thick disk based on the large column densities,
particle densities, and masses of the clouds.

This conclusion was motivated in part by a detailed comparison of the extraplanar dust
filaments and DIG emission by \pIII, which demonstrated that they were tracing independent
structures in the thick disk ISM of \ngc 891.\footnote{Keppel et al. (1991) reached the
same conclusion on the basis of densities estimated for the DIG and extraplanar dust
features.}  Following the study of \ngc 891, Thompson \etal\ (2004; hereafter THS04), used
the {\em Hubble Space Telescope} (\hst) to investigate the extraplanar dust in the
edge-on galaxy, \ngc 4217. These authors indicate that in addition to the smoother
distribution of the \Ha\ emission morphology compared to the extraplanar dust structures,
they find little spatial correlation between areas of enhanced \Ha\ emission around
regions of concentrated OB associations and high-\z\ dust structures. Rossa \etal\ (2004)
similarly used images from (\hst) to show the ISM traced by extraplanar dust was distinct
in morphology from the DIG in \ngc 891, while Rossa \& Dettmar (2003a, 2003b) argued for a
similar arrangement in their larger sample of galaxies. The dusty extraplanar filaments
are physically distinct from the DIG, even though the presence of these two tracers of the
extraplanar ISM is strongly correlated (Rossa \& Dettmar 2003b; HS99).  The small scale
structures in the \pIII\ and subsequent Rossa \etal\ (2004) imaging, the high column
densities ($\Nh \ga 10^{21}$ in many cases), large cloud masses, and high particle
densities ($n_{\rm H} \ga 1-25$ \percc ; \pIII, Howk 2005) are all features in great
contrast to the other observed phases of the thick disk ISM and point to a denser, cooler
medium than the DIG. Thus, \pIII\ concluded that all of the thermal phases of the ISM
found in the thin disks of spiral galaxies were also present in their thick disks.

The only galaxy for which a detailed comparison of the \Ha\ (DIG) and extraplanar dust
morphologies has been undertaken is \ngc 891. While the structures traced by these two
probes of the extraplanar ISM were almost completely distinct in \ngc 891, the
universality of the result is not settled.  In particular, \Ha\ images of the DIG are
often taken at relatively low spatial resolution, in part due to the faintness of the
emitting gas, while the dust structures are lost to confusion at similar resolution. \ngc
891 has one of the brightest known \Ha -emitting DIG layers (e.g., Rand 1996, Miller \&
Veilleux 2003) and has more prominent filaments within the DIG than many galaxies (Heald
\etal\ 2007).  For galaxies with lower star formation rates, where the brightness of the
DIG and the prominence of filaments diminishes, are there corresponding changes in the
extraplanar dust?  In this work we present high-resolution images of extraplanar dust and
DIG in the edge-on galaxies \ngc 4013 and \ngc 4302.  Our broadband images include
observations from \hst\ that elucidate the smallest scale structures in the dusty thick
disk, while our \Ha\ images are among the highest spatial resolution acquired from the
ground and have good continuum subtraction.  We use these images to extend the detailed
comparison of extraplanar dust and DIG properties to these two galaxies, which have
fainter DIG emission and a smaller fraction of the emission arising in filaments (i.e., a
larger proportion coming from the smooth DIG component). In the end we find no strong
physical connection between the DIG and the structures traced by high-contrast dust
absorption in these galaxies, similar to what is seen in \ngc 891. The DIG and dust-laden
clouds trace fundamentally different material in the thick disk.

Our work is organized as follows.  We discuss the observations and data reduction in \S
\ref{sec:observations}.  In \S \ref{sec:thickdisk} we discuss the properties of the
extraplanar dust and associated gas, while in \S \ref{sec:dig} we describe the properties
of the extraplanar \Ha\ emission (DIG).  In \S \ref{sec:comparison} we provide a direct
comparison of the extraplanar dust morphologies and the DIG detected in \Ha.  We discuss
in \S \ref{sec:discussion} the origins, nature, and properties of the multiphase ISM in
both \ngc 4013 and \ngc 4302 and summarize the major results of our work in \S
\ref{sec:summary}.

\begin{figure*}[!ht]
\begin{center}
  \includegraphics*[scale=0.9]{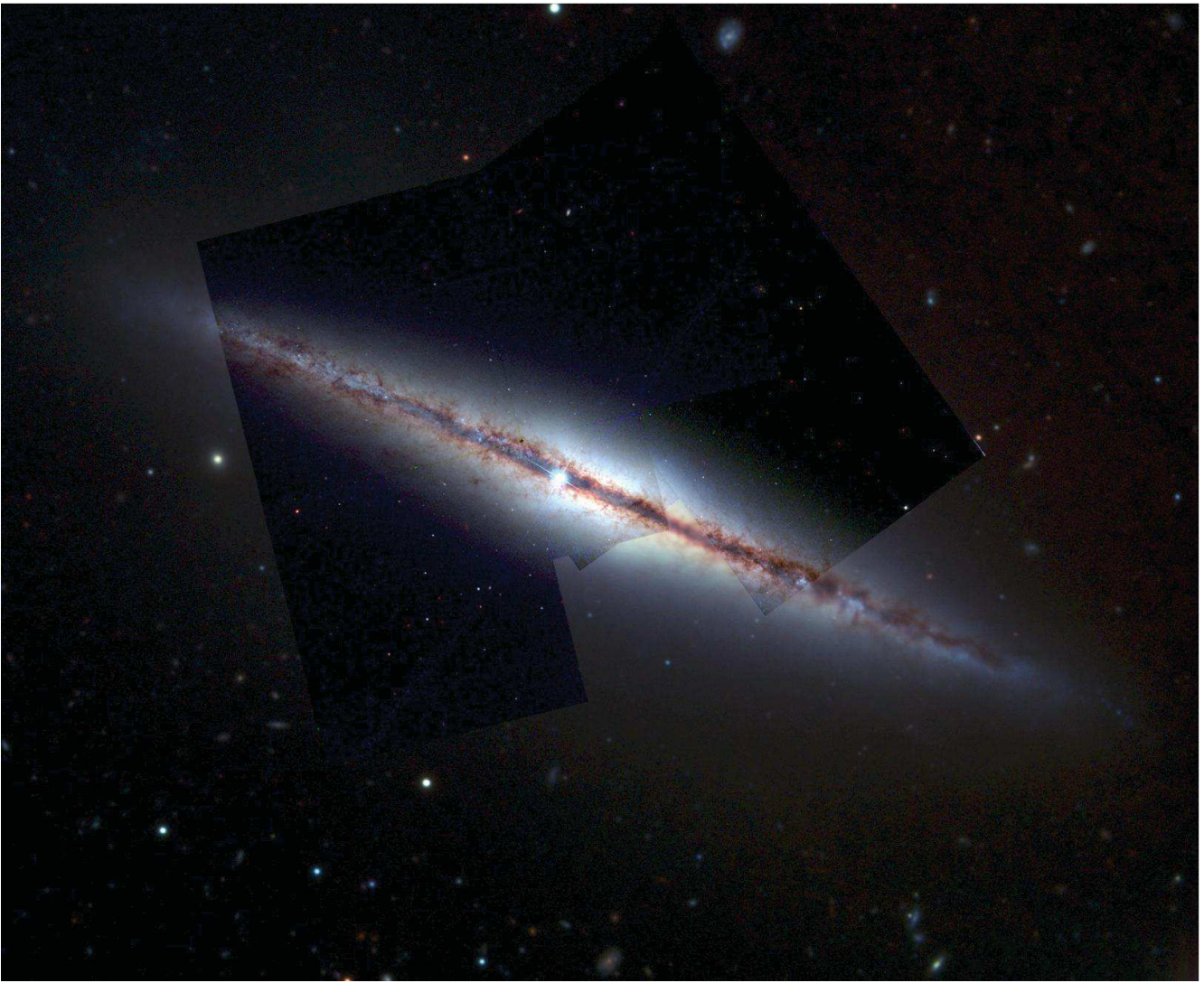}
  \caption{Three color (BVI) composite mosaic  of \ngc 4013 created from the \hst\ and
  WIYN imaging of this galaxy.  The field of view of this image is $\sim 5\arcmin \times
  5\arcmin$ on a side.  North is up, east is to the left. Dusty thick disk clouds are seen
  as dark patches of extinction off of the galactic plane.  The \hst/WFPC2 images of this
  galaxy cover the central and northeastern section of the galaxy but do not include much
  of the southwestern part of the galaxy.  The bright emission in the plane just northeast
  of the bulge is a foreground star.  There are slight edge artifacts due to the merging
  of the \hst\ and WIYN images.}
  \label{fig-n4013color}
\end{center}
\end{figure*}

\section{Observations and Reductions}
\label{sec:observations}
The work in this paper is based on optical imaging of two galaxies -- \ngc 4013 and \ngc
4302 -- whose properties are listed in Table \ref{table:galaxies}.  We use broadband and
narrow-band imaging from several sources: 1) \hst/Wide Field Planetary Camera 2 (WFPC2))
images of both galaxies in the (WFPC2) equivalents of the BVI bands to study extraplanar
dust absorption at the highest resolution; 2) ground-based broadband images of NGC 4013
and NGC 4302 taken with the WIYN 3.5-m and LBT 2$\times$8.4-m telescopes, respectively, to
study extraplanar dust at larger radial and vertical distances from the galaxy centers
than allowed by \hst; and 3) narrow-band \halpha\ imaging of the extraplanar DIG in these
galaxies, taken with the WIYN 3.5-m, to study the connection (or lack thereof) between the
ionized gas and material traced by dust absorption in the thick disks of these galaxies.
We discuss the processing of these datasets in this section.

\vspace{20pt}
\subsection{{\em HST}/WFPC2 Observations and Processing}

We obtained broadband imaging observations of \ngc 4013 and \ngc 4302 with the Wide Field
Planetary Camera 2 (WFPC2) aboard \hst. WFPC2, described fully in Biretta \etal\ (2002),
consists of four cameras. The three Wide Field Cameras (WFCs) together cover an
``L''-shaped region 150\arcsec $\times$150\arcsec\ with intrinsic spatial sampling of
$0\farcs1$ per pixel (f/12.9).  The Planetary Camera (PC) covers a 34\arcsec
$\times$34\arcsec\ square field with $0\farcs046$ per pixel (f/28.3).

Our observations, summarized in Table \ref{table:hstlog}, were acquired as part of GO
program 8242 (PI: Savage). We imaged the majority of the galaxy \ngc 4302 in one
pointing, while we used two separate pointings to cover a significant fraction of \ngc
4013.  For \ngc 4302 and one of the \ngc 4013 pointings, we acquired eight exposures in
the F450W filter and four exposures in each of the F555W and F814W filters. We obtained
four images in each of the three filters for the secondary \ngc 4013 pointing.  The F450W,
F555W, and F814W filters roughly correspond to the Johnson-Cousins B, V, and I
bands\footnote{We chose the F450W filter as the B-band equivalent because it has a higher
over-all throughput than the F439W filter; the transformation to the standard
Johnson-Cousins system is well-behaved (see Holtzman \etal\  1995).}.

The images acquired for each pointing were split equally between two dithered
sub-positions, with a linear dither step of 3.5 WF CCD pixels between the positions.  We
used the Dither II package and its drizzle algorithms (see Fruchter \etal\  1998 and
Fruchter {\&} Hook 1997) to combine the dithered images onto a finer grid, following
THS04. The resulting cosmic ray-cleaned images have an angular sampling of $0\farcs05$ per
pixel.

\begin{figure*}[!ht]
\begin{center}
  \includegraphics*[scale=0.9]{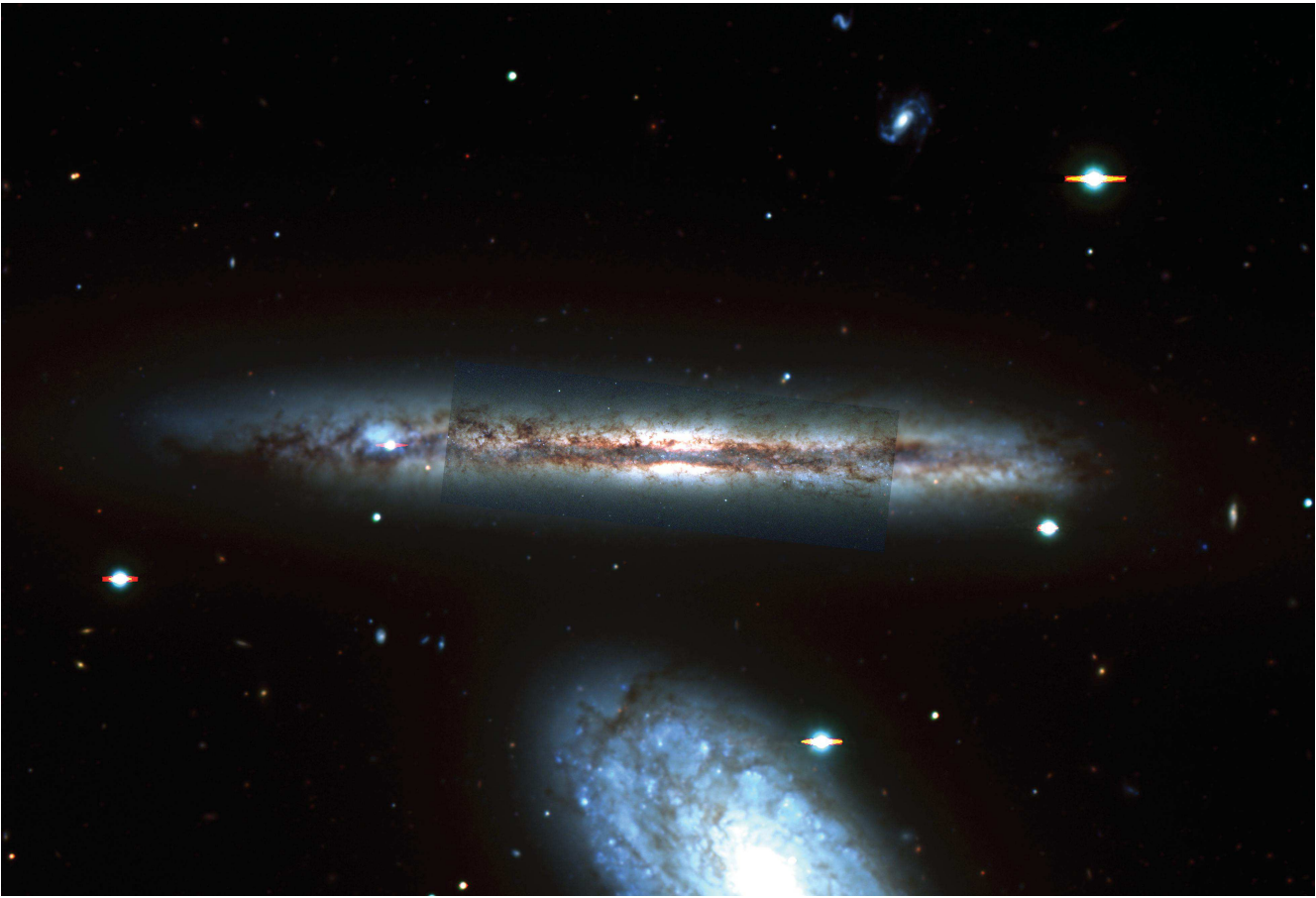}
  \caption{Three color (BVI) composite mosaic image of \ngc 4302 created from the \hst\
  and LBT imaging of this galaxy.  The field of view of this image is $\sim 8\arcmin
  \times 5\arcmin$ on a side. North is to the right, east is up. Dusty thick disk clouds
  are seen as dark patches of extinction off of the galactic plane.  The \hst/WFPC2 images
  of this galaxy cover the central region out to R $\sim 6$ kpc. \ngc 4302 shows no
  signature of interaction with the neighboring galaxy, \ngc 4298, seen in the bottom of
  the composite image. There are slight edge artifacts due to the merging of the \hst\ and
  LBT images.} 
  \label{fig-n4302color}
\end{center}
\end{figure*}

Figures \ref{fig-n4013color} and \ref{fig-n4302color} show composite color images of \ngc
4013 and \ngc 4302, respectively, created from the WFPC2 BVI images.  The WFPC2 images
have been combined with ground-based observations in these images.  The ground-based
images are more sensitive to faint, extended structures (i.e., the faint thick disk
stellar emission) and probe a larger field of view.  For NGC 4013, the ground-based data
are from the WIYN 3.5-m telescope and are discussed in \S\ref{sec-wiyn}.  The ground-based
observations of NGC 4302 were obtained with the 2$\times$8.4-m Large Binocular Telescope
(LBT) and are described in \S\ref{sec-lbt}.

\subsection{WIYN Observations and Processing}
 \label{sec-wiyn}
To study the distribution of the diffuse ionized gas in these galaxies, we acquired
narrow-band \halpha\ and Gunn \sdssr\ images of \ngc 4013 and \ngc 4302 at the WIYN 3.5-m
telescope at Kitt Peak National Observatory.  These data were collected on 2004 April 18
and 19 (UT) under non-photometric conditions using the WIYN Tip-Tilt Module (WTTM),
although without the tip-tilt functionality enabled.  A log summarizing the WIYN
observations is given in Table \ref{table:wiynlog}, including the total exposure time in
each filter and the seeing-limited resolution in the final images for each galaxy.

The WTTM \footnote{The WTTM instument described here: {\tt
http://www.noao.edu/wiyn/WTTM\_manual.html}.}  is an optical re-imaging system working at
f/7.43.  This system feeds a $2048\times4096$ EEV CCD with $13.5$ $\mu$m pixels.  Each
pixel subtends $0\farcs1125$ on the sky giving a $3\farcm8 \times 4\farcm7$ field of view
that is vignetted at the corners.  The data acquisition, calibration, and reduction of the
images were done in a manner similar to that described in \pIII. We note that because the
field of view of the imager is relatively small compared with the galaxies themselves, the
sky background subtraction is somewhat uncertain.  None of our conclusions are sensitive
to this problem; however, it precludes us from, e.g., properly deriving scale heights of
the diffuse ionized gas in these galaxies.

We derived the astrometric plate solution for our images using a grid of stars whose 
coordinates were measured in the Digitized Sky Survey for the \ngc 4302 observations. 
These agree well with the \hst -provided plate solution for the \wfpc\ data.  The rms
error in using the plate solution to derive coordinates is always significantly less than
1\farcs0.  For the \ngc 4013 observations, this treatment of the data left a systematic
offset between the WIYN and \hst\ data.  Due to a lack of DSS-detected stars in the small
\wfpc\ field, we used stars and galaxies in common between the \wfpc\ and WIYN images to
bring the WIYN data onto the same reference frame as the \hst\ data.  The BVI imaging of
\ngc 4013 was obtained with the previous generation WIYN Imager, the S2KB Camera.  The
properties and reductions of these data are as those discussed in \pIII. These data are
also summarized in Table \ref{table:wiynlog}.

The principal purpose of our ground-based WIYN observations is to study the distribution
of \halpha\ emission from the DIG in these galaxies.  The WIYN W15 filter was used as an
on-band \Ha\ filter.  This filter, centered at $6569$\AA\ with a FWHM of $\sim73$\AA,
contains emission from the \Ha\ and nearby [\ion{N}{2}]\footnote{While the images contain
both \Ha\ and [N$\,$II] emission, for brevity's sake we will hereafter refer to our images
as \Ha\ images.  The reader should be aware that the bandpass also contains the [N$\,$II]
transitions.} lines plus a stellar continuum contribution. We used the Gunn \sdssr\
observations to estimate the stellar continuum and subtract it from the narrow-band
images, scaling the \sdssr\-band images so that faint Milky Way foreground stars were
completely removed from the \Ha\ images after continuum subtraction. Non-saturated
foreground stars and background galaxies are well subtracted in our final line images.  We
found no evidence for a need to adopt a different scale factor for the bulge of the
galaxies compared with the outer disks of the galaxies (e.g., Rand 1996). The Gunn \sdssr\
filter used for the continuum estimate encompasses the \halpha\ line itself and may lead
to slight over-subtraction of the line emission. This has little impact on the current
paper, in which we limit ourselves to morphological considerations.

\subsection{LBT Observations and Processing}
\label{sec-lbt}
We obtained broadband UBVI images of NGC 4302 with the Large Binocular Cameras (LBCs) on
the 2$\times$8.4-m LBT under non-photometric conditions on 2008 May 03.  A log summarizing
the LBT observations is given in Table \ref{table:lbtlog} including the total exposure
time in each filter and the seeing-limited resolution in the final images for \ngc 4302. 
The LBT uses two co-pointed 8.4-m mirrors to feed a pair of nearly-identical LBCs.  The
LBCs are described in Giallongo \etal\ (2008).  The two cameras operate simultaneously at
prime focus, one optimized for blue wavelengths and the other for red (both can provide
V-band imaging, the approximate cross-over in sensitivity).  Each camera provides a
$\sim23\arcmin \times 23\arcmin$ field of view using a four CCD mosaic with $0\farcs23$
pixels. Inter-chip spacings are filled in by dithering the telescope over several
exposures.  A log of our observations is given in Table \ref{table:lbtlog}.  We used the
LBC-Blue camera to image NGC 4302 in the U and B filters, while the LBC-Red camera
provided coverage of in the V and I filters.  We use twilight sky exposures to perform the
flat field corrections.

Reduction of the LBC images requires combining a series of dithered images, taking into
account the distortions that occur over the very wide field of view. We largely follow the
techniques described in Sand \etal\ (2009) using modified versions of scripts from B.
Weiner et al. (in preparation).  The data are processed in two steps.  The first uses
standard reductions from the IRAF package {\em mscred} to trim and remove the overscan
region, subtract bias frames, and apply flat fields (derived from twilight flats). 
Subsequently, we used the SCAMP (Bertin 2002) and SWARP (Bertin \etal\ 2002) software
packages\footnote{{\tt http://www.astromatic.net/}} to produce the final images.  SCAMP
was used to derive the astrometric solution for each of the chips in each image.  The
final solution is based on cross-matching sources identified in the LBT images using
SExtractor (Bertin \& Arnouts 1996) with an astrometric catalog derived from DR7 of the
Sloan Digital Sky Survey (SDSS-DR7; Abazajian \etal\ 2009). This is a necessary step due
to the distortions over the full field given the $f/1.14$ focal ratio.  The absolute
astrometric calibration has an rms dispersion compared with SDSS of $\sim0\farcs2$ across
the entire field of view.  The individual images were subsequently resampled and coadded
using SWARP.

\section{The Dusty Thick Disks of \ngc 4013 and \ngc4302}
\label{sec:thickdisk}
Visual inspection of Figures \ref{fig-n4013color} and \ref{fig-n4302color} reveals the
presence of strong, patchy extinction in the thick disks of these two galaxies.  Filaments
and clouds of absorbing material are seen to nearly the radial edge of the galaxies'
optical disks and to \z -heights approaching the largest distances to which stellar light
can be seen in the relatively shallow \wfpc\ images, although the ground-based LBT and
WIYN images reveal the dust extinction extends to lesser heights than the stellar thick
disk. This extraplanar extinction is a direct tracer of the thick disk ISM in these
galaxies (e.g., Howk 2005).  In order to be seen, these dusty regions must have higher
column densities than their surroundings with these filaments having excess extinction
compared with nearby sight lines, providing significant contrast between these clouds and
their local environment in our images. Smoothly distributed dust does not produce such
contrast; thus a diffusely distributed component of dust may exist undetected in the thick
disks of these galaxies. That the column densities of the observed dust-bearing clouds are
higher than their surroundings almost certainly implies the dust features seen in our
images have significantly higher volume densities than their surroundings, as well.

\begin{figure*}[!ht]
\begin{center}
  \includegraphics*[scale=0.8]{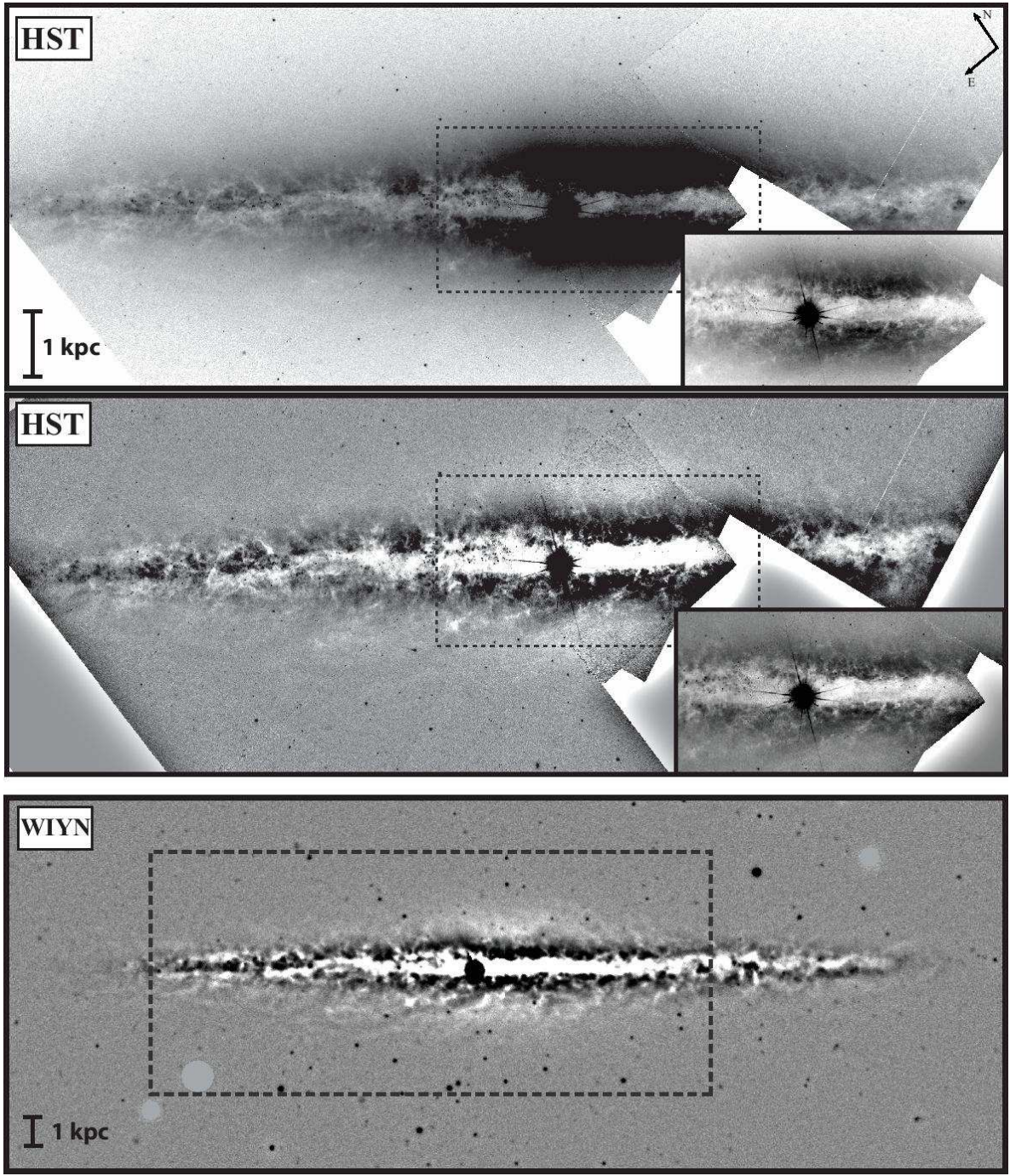}
  \caption{\hst\ \& WIYN V-band images of \ngc 4013. The top panel shows the \hst\ V-band
  image; the middle panel shows the unsharp masked version of this image. The insets in
  the top two panels are stretched to show the bulge region in greater detail. The area of
  the insets is indicated by dashed lines. The bottom panel shows the unsharp masked
  version of the WIYN V-band image. The dashed box in the WIYN V-band indicates the extent
  of the \hst\ image shown in the upper panels. All image displays are inverted so that
  areas of dust extinction appear lighter than their surroundings. The \hst\ images show
  dense, narrow, filamentary-like extraplanar dust structures at high resolution.  The
  WIYN image show the extensive dense dust structures and complexes extending to heights
  $z \sim2$ kpc beyond the midplane.}
  \label{f-ngc4013v}
\end{center}
\end{figure*}

\begin{figure*}[!ht]
\begin{center}
  \includegraphics*[scale=0.8]{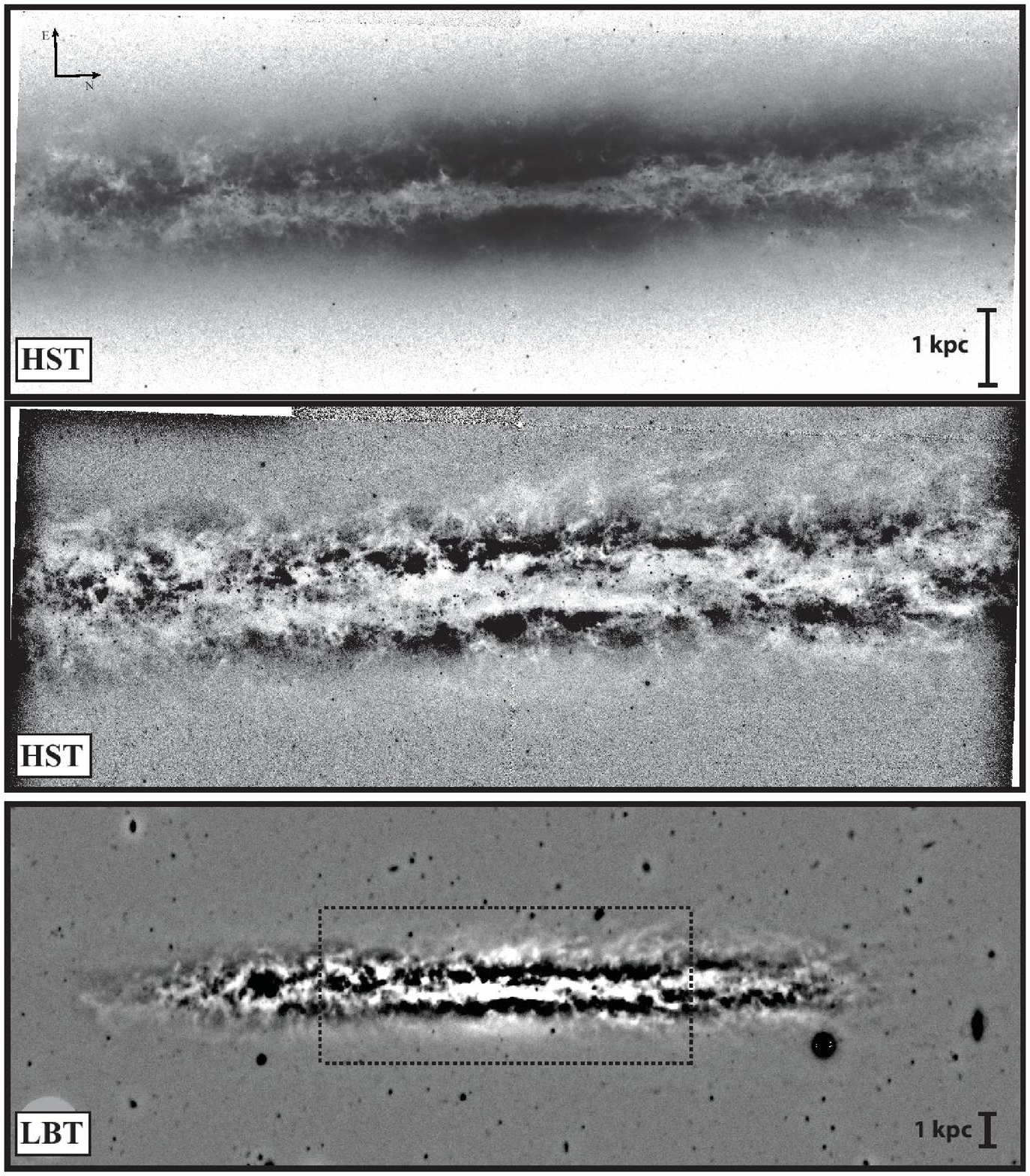}
  \caption{\hst\ \& LBT V-band images of \ngc 4302. The top panel shows the \hst\ V-band
  image; the middle panel shows the unsharp masked version of this image.  The bottom
  panel shows the unsharp masked version of the LBT V-band image.  The dashed box in the
  LBT V-band indicates the extend of the \hst\ image shown in the upper panels. All image
  displays are inverted so that areas of dust extinction appear lighter than their
  surroundings. The \hst\ images show a large number of extraplanar dust structures at
  high resolution. The LBT image shows extraplanar dust-bearing clouds in the thick disk
  along nearly the entire radial length of the galaxy and is more sensitive to small
  extinction clouds.}  
  \label{f-ngc4302v}
  \end{center}
\end{figure*}

Figure \ref{f-ngc4013v} shows the \hst\ and WIYN V-band images of \ngc 4013. The top panel
shows the \hst\ V-band data and the middle panel shows the unsharp masked version of the
top panel. The display of the galaxy in the top panel is not able to show the absorbing
clouds against the bulge and further out in a single display, making the unsharp masking
necessary. The unsharp masked versions of our images remove the large-scale gradients in
the underlying stellar light, revealing structure on scales smaller than the smoothing
kernel (FWHM$\, \la 350$ pc) (see \pIII\ and \ths\ for detailed discussions of this
approach).  We only use these images for display and have taken care to assess and
minimize masking-related artifacts. Both \hst\ images include insets showing the central
bulge region with a stretch that shows the dust absorption more clearly. The bottom panel
shows the unsharp masked display of our final WIYN data. The
WIYN image gives almost full radial coverage of the galaxy, revealing extraplanar dust
structures extending along nearly the entire radial length of the disk of \ngc 4013.
Extraplanar dust clouds can be identified against the background stellar light $\z\ \sim
2.0$ kpc from the midplane. Those dust clouds at the largest heights show less prominent
substructure and generally have lower $a_{V}$ values, but they are still well detected in
our WIYN images.  We see no evidence for disparities in the amount of extraplanar dust
comparing opposite sides of the plane. 

Similarly, Figure \ref{f-ngc4302v} shows the \hst\ and LBT V-band images for \ngc 4302.
The top panel shows the \hst\ V-band data and the middle panel shows the unsharp masked
version of the top panel. The \hst\ images cover $\sim 6$ kpc projected radial distance
from the center on either side, while the LBT images show $\sim 13$ kpc radial span of the
optical disk and more. The dashed box in the LBT V-band indicates the extend of the \hst\
image shown in the upper panels. Extraplanar dust clouds can be identified against the
background stellar light $\z\ \sim 2.5$ kpc from the midplane, detected in the LBT images.
The FHWM of the Gaussian used to produce the unsharp masked images for \ngc 4302 were the
same as in \ngc 4013.

\vspace{10pt}
\subsection{Quantifying Dust Extinction}

Using the \hst\ V-band data for our measurements (top panels of Figure \ref{f-ngc4013v}
and \ref{f-ngc4302v}) and following \pIII\ and \ths, we provide spatially-averaged
estimates of the apparent opacities of a small number of extraplanar dust structures. 
From these we derive the ``apparent'' extinction, column densities, particle
density, and masses of these dust-bearing structures.
We define the ``apparent'' extinction, $a_{\lambda
}$, for a given wave band $\lambda $ as
\begin{equation}
\label{eq1}
a_\lambda =-2.5 \log (S_{dc,\lambda} / S_{bg,\lambda} );
\end{equation}
here $S_{dc,\lambda }$ is the average surface brightness measured toward a dust cloud
(dc), and $S_{bg,\lambda }$ is the surface brightness of the local background (bg).
Defined in this way, the apparent extinction is a lower limit to the true extinction,
$A_{\lambda }$, because $S_{dc,\lambda }$ includes both extincted starlight from behind as
well as unextincted light emitted in front of the feature.  One can, in theory, use
$a_{\lambda}$ for multiple wavelengths with an assumed extinction curve to estimate the
fraction of light originating in front of the dust feature (e.g., Gallagher {\&} Hunter
1981, Knapen \etal\ 1991, HS97).  In this case, the fraction of $S_{dc,\lambda }$ arising
in front of the clouds is $x$ and the true extinction is a function of $a_{V}$ and $x$.
However, \pIII\ found that this approach does not always lead to self-consistent results,
likely due to the effects of spatial averaging of probing differing depths into clouds at
different wavebands and of scattering. \pI\ (who used a narrower range of wavelengths)
provided evidence for a significant amount of stellar light arising in front of the dust
clouds in \ngc 891.

Here we follow \pIII\ and \ths\ in assuming $A_V \ga a_V$ (equivalent to assuming $x \ge
0$).  This has the advantage that our calculations of the physical properties of the
structures seen in our images will provide lower limits to the true properties.  \pIII\
discuss the degree to which $A_V$ departs from this assumption for several values of $x$.
For example, for $x=0.25$, the V-band extinction $A_V \approx 1.5 \, a_v$. We do not
report the extinction in the other bands.

For each dust feature we measure the ratio of surface brightnesses $S_{dc,\lambda }/
S_{bg,\lambda }$ using intensity distributions measured vertically or horizontally, with
respect to the disk, averaging the results for a given cloud. We estimated the background
surface brightness appropriate for each feature by fitting a spline to the intensity
distribution of the star light in each cut. This technique is different than that used in
our previous works where we were more concerned with the impact of the smearing due to the
seeing.  The current measurements will average the $a_{V}$ values over more of the clouds
than those earlier measurements, leading to somewhat smaller extinction values.

\subsection{Physical Properties of Individual Dust Structures}

\begin{figure*}[!ht]
\begin{center}
  \includegraphics*[scale=.9]{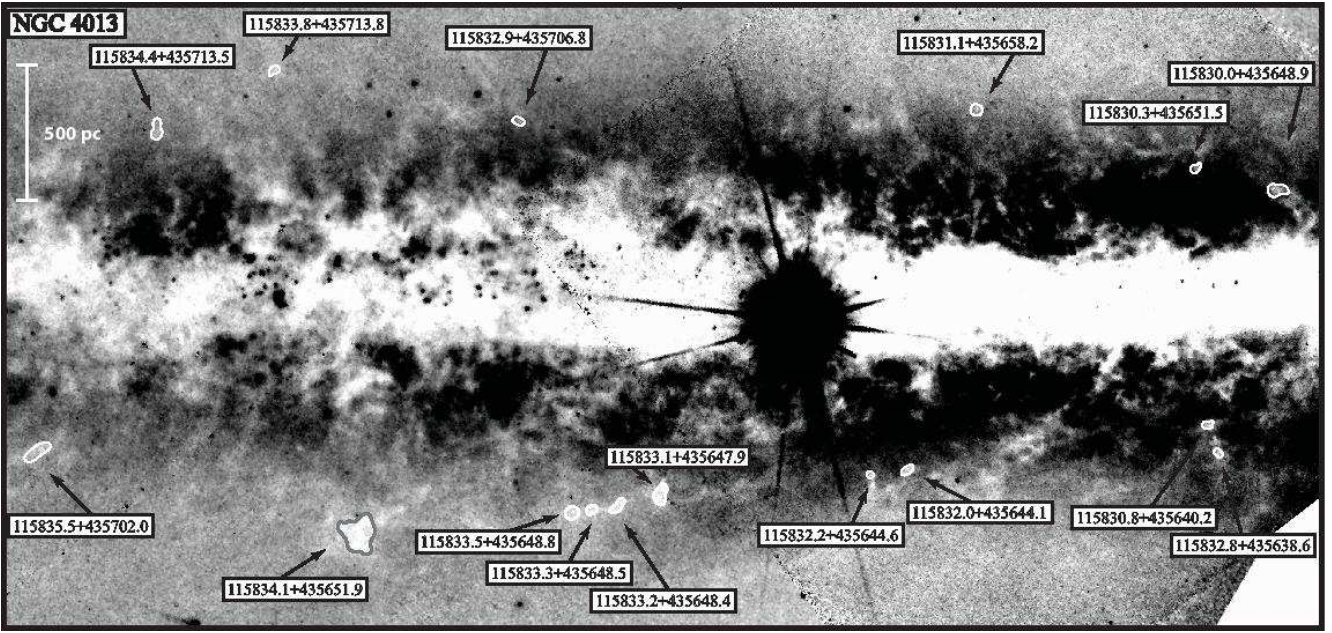}
  \caption{\hst\ unsharp masked V-band image of \ngc 4013 with select dense extraplanar
  clouds outlined and labeled.  The properties of these clouds are summarized in Table
  \ref{table:dustproperties}. These particular absorbing structures were chosen based on
  their location, apparent opacity, and size in an effort to emphasize the physical
  properties of smaller, dense structures that the high-resolution \hst\ images reveal.
  These extraplanar clouds are located at heights $\z\ \sim 430 - 930$ pc, have estimated
  column densities and masses $N_{\rm H}>2\times\e{20}$ cm$^{-2}$ , and $M \ga \e{4} $
  $M_\odot$. Substructure in these clouds is seen to the limit of our resolution. The
  orientation of this image is as in Figure \ref{f-ngc4013v}.  Figure
  \ref{f-ngc4013-clump} gives a close-up image of the gray outlined clump to the east
  below the bulge, cloud $115834.4+435651.9$.}
  \label{f-ngc4013_clouds}
\end{center}
\end{figure*}

\begin{figure*}[!ht]
\begin{center}
  \includegraphics*[scale=0.9]{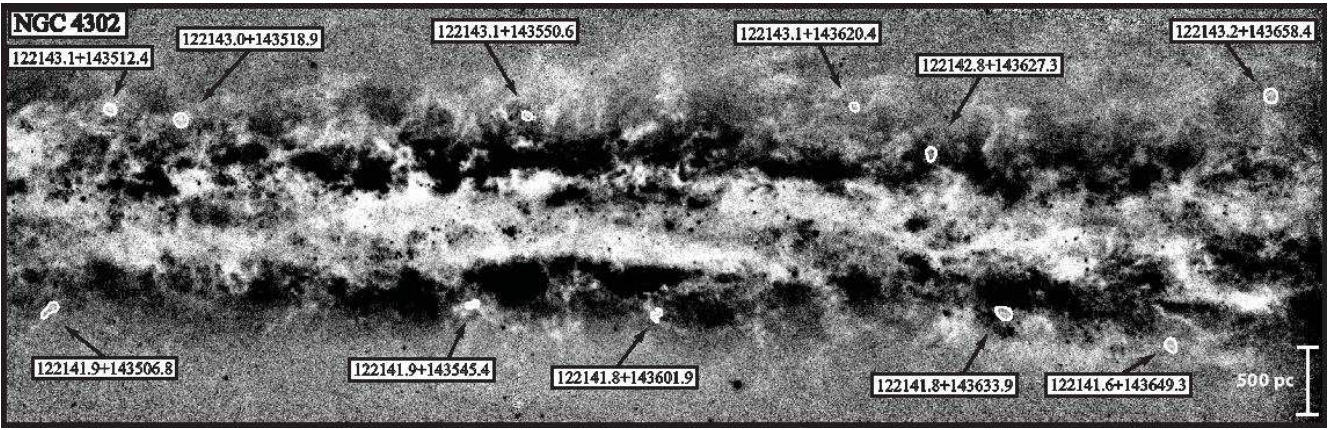}  
  \caption{\hst\ unsharp masked V-band image of \ngc 4302 with select dense extraplanar
  clouds outlined. The orientation of this image is as in Figure \ref{fig-n4302color}. The
  properties of these clouds are summarized in Table \ref{table:dustproperties}. These
  particular absorbing structures were chosen based on their location, apparent opacity,
  and size in an effort to emphasize the physical properties of smaller, dense structures
  that the high-resolution \hst\ images reveal. These extraplanar clouds, located at
  heights $\z\ > 590$ pc, have estimated column densities and masses $N_{\rm H} >
  2$$\times$$\e{20}$ cm$^{-2}$, and $M>$$\e{4} $ $M_\odot$, respectively. Substructure in
  these clouds is seen to the limit of our resolution.}
  \label{f-ngc4302_clouds}
\end{center}
\end{figure*}

Table \ref{table:dustproperties} summarizes the physical properties of several cloud
complexes in the thick disks of the two target galaxies from the \hst\ V-band data.
Figures \ref{f-ngc4013_clouds} and \ref{f-ngc4302_clouds} identify the clouds discussed in
Table \ref{table:dustproperties}, they are displayed in the unsharp-masked \wfpc\ images
to better show the small-scale features of the selected clouds. Several measured
quantities are given, including their positions, physical dimensions, height above the
midplane, and $a_V$ values.  Estimates of the column densities, densities, and masses
associated with the clouds are also given, as outlined below.  In this work we have
favored small, relatively isolated structures in order to study the densities implied by
the small-scale structure observed in the \wfpc\ imaging.  The choice to work with
isolated clouds tends to favor lower extinction clouds at relatively large \z.  The
smaller sizes (to emphasize high densities) tend to favor lower masses.

The values of $a_V$ in Table \ref{table:dustproperties} are in the range 0.1--0.3 mag.
These values represent spatial regions of the clouds lower than the values appropriate for
the cores of the structures. Given the values of $a_{V}$ are also lower than the true
extinction, realistic values of $a_{V}$ for much of the clouds are likely significantly
larger than those values given in Table \ref{table:dustproperties}.

Transforming the apparent extinctions discussed above into physical properties of the
clouds requires an assumption about the gas-to-dust relationship in the thick disk.
Following earlier works, we assume the clouds follow dust to gas relationships appropriate
for the disk of the Milky Way (\pI, \pII, \pIII).  This is a necessary step for estimating
the physical conditions of the dust features from the measured apparent extinctions, one
whose consequences are discussed at length in \pIII. We adopt the average relationship
between total hydrogen and the color excess $E(B-V)$ of diffuse clouds in the Galactic
disk as estimated by Bohlin \etal\ (1978): $N(\mbox{\HI} + {\rm H}_{2})/E(B-V) = 5.8
\times 10^{21}$ atoms cm$^{-2}$ mag$^{-1}$.  To relate this to the extinction, we assume
$R_{V }\equiv A_{V }/ E(B-V) \approx 3.1$, roughly the mean for diffuse clouds in the
Galactic disk (e.g., Cardelli \etal\ 1989).  This gives the total hydrogen (\HI\ + \htwo)
column density from the apparent extinction:
\begin{equation}
\Nh ({\rm cm}^{-2}) > 1.9 \times  10^{21}  a_{V}.
\end{equation}
The inequality arises since $a_V < A_V$.  Since both \ngc 4013 and \ngc 4302 are
relatively massive spiral galaxies, the assumption of a Milky Way gas-to-dust ratio will
probably yield a reliable estimate of the associated gas column density.  However, if
significant dust destruction has occurred in any particular feature, the derived \Nh\ will
be an underestimate, while if the dust has been separated from the gas then \Nh\ will be
an overestimate.  We favor scenarios in which little dust destruction occurs in moving the
dust from the thin to thick disk or where some destruction has occurred in which the dust
is separated from the gas.  We thus quote the column densities as lower limits.

We give lower limits to the masses of for the features in Table
\ref{table:dustproperties}, calculated as $M \sim \mu m_H \Nh A_{dc}$, where $A_{dc}$ is
the projected area of the dusty clouds.  We assume $\mu = 1.37$ to correct for the
contribution to the mass from helium and the heavy elements.  The definition of the cloud
areas is a critical and highly-subjective component of the mass estimates.  In the present
work we have favored small clouds and have been conservative in our derivation of the
projected areas. The implied masses should be treated as lower limits.

In addition to the columns and masses, we also include a very crude density estimate in
Table \ref{table:dustproperties}.  This estimate assumes the clouds we see are largely
cylindrically-symmetric such that $n_{\rm H} \sim \Nh / \Delta x$, where $\Delta x$ is the
width of the minor axis of the clouds.  The only support for this assumption is the
general prevalence of filamentary absorbing structures seen in our images, although this
impression could be affected by the general confusion caused by overlapping clouds along
the line of sight.  In general the morphologies of the absorbing clouds are complex, but
we move forward using this simplistic assumption for estimating the densities.

Keeping in mind the uncertainties discussed above, the properties of the clouds summarized
in Table \ref{table:dustproperties} are nonetheless intruiging.  The features all have
estimated columns $\Nh \ga 2 \times 10^{20}$ cm$^{-2}$, simply as a matter of selection
(it is difficult to identify clouds with $a_v < 0.1$).  We infer cloud masses $\ga 10^4
M_{\odot}$ for all but the smallest two clouds in Table \ref{table:dustproperties}, and
the densities implied for those clouds included in the table are of order $n_{\rm H} \sim
1-4$ \percc.  Additionally, we examined some of the darkest narrow filaments in the \wfpc\
images located in regions too confused to show well in Figures \ref{f-ngc4013_clouds} and
\ref{f-ngc4302_clouds}, so they are not included in Table \ref{table:dustproperties}.
Several of these clouds have column densities approaching or exceeding $10^{21}$
cm$^{-2}$, giving mean densities $n_{\rm H} \ga 8-10$ \percc, which is extraordinary for
thick disk clouds.
\begin{figure*}[!ht]
\begin{center}
  \includegraphics*[scale=.9]{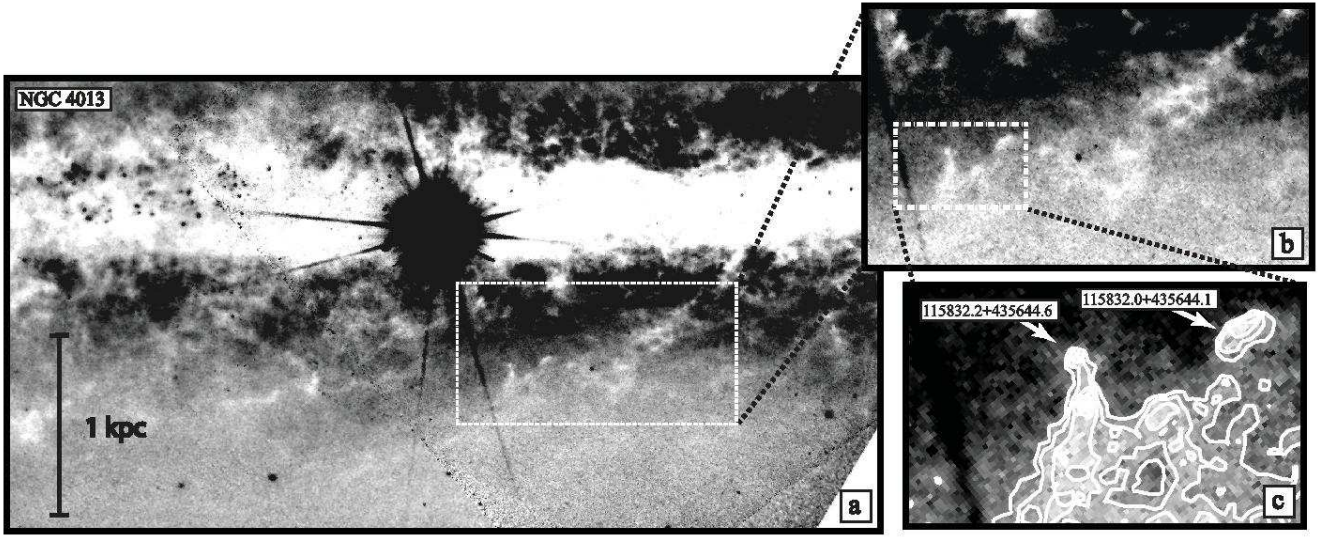}
  \caption{(a) A section of the bulge of \ngc 4013 from the \hst\
    unsharp masked V-band images showing the large dust complex
    discussed in the text. (b) Close-up view of the large complex,
    which extends $\sim900$ pc by 200 pc.  This complex has a mean
    extinction $a_{V} = 0.04$ and total mass $\ga 10^{5}
    M_{\odot}$. (c) An enlarged view of two of the cometary-like
    clouds associated with the large complex. These are clouds
    $115832.2+435644.6$ and $115832.0+435644.1$ in Table
    \ref{table:dustproperties}. Cloud $115832.2+435644.6$ has a
    notable core-halo type structure, as emphasized in the relative
    contours shown on this display. The core of the cloud has $a_{V} =
    0.36$ and implied density $\nH \geq 7$ cm$^{-3}$. The orientation
    of these images is as in Figure \ref{f-ngc4013v}.} 
  \label{f-ngc4013-cometary}
\end{center}
\end{figure*}
The column densities and densities measured for the thick disk clouds in \ngc 4013 and
\ngc 4302 are similar to those seen in \ngc 891 and \ngc 4217 (\pI, \pIII, \ths, Howk
2005).  The derived densities likely depend on the resolution with which one observes the
structures, as higher resolution images not only smear less background light into the
structure but also allow the identification of smaller-scale filaments. The masses given
in Table \ref{table:dustproperties} are smaller than have been reported in earlier works
(\pII\, \pIII, \ths), but that is largely a function of our newly adopted measurement
technique and the choices we have made for emphasizing the smaller subcomponents rather
than the largest complexes.  Complexes of absorption with implied total masses well in
excess of $10^5 M_{\odot}$ certainly exist in these galaxies.  Overall, we do not see any
clear evidence that the dust-bearing thick disk clouds in these galaxies are significantly
different in character than those seen in \ngc 891 (\pIII) or \ngc 4217 (\ths).

\subsection{Apparent morphologies of dust-bearing gas clouds}

Our images allow us to identify large-scale dust-bearing clouds in the thick disks of
these galaxies and study their internal structure.  Our high angular resolution \hst\
images reveal substructure within the cloud complexes, down to the limit of our
resolution, but only for the inner sections of both galaxies. Our LBT and WIYN data allow
us to identify dust clouds at larger radial and vertical distances due to their greater
field of view and depth compared with the \hst\ images. Similar to clouds in the Milky
Way, in which structure is seen on many scales, the thick disks of \ngc 4013 and \ngc 4302
contain narrow filamentary structures, large complexes of clouds (specifically in \ngc
4302), as well as small, very dense clouds seen to our resolution limit.

\subsubsection{\ngc 4013}

The extraplanar dust-bearing clouds in \ngc 4013 are generally highly structured
filaments, although we see a broad range of morphologies in both the \hst\ and WIYN images
in Figure \ref{f-ngc4013v}. The typical dust filaments are $ \sim 100$ pc by a few $\times
100 $ pc in size (see Table \ref{table:dustproperties} and Figure \ref{f-ngc4013v}). We
note that the dust filaments are unlikely to be associated with a flare in the outer
galaxy (see detailed discussion in \pII, \pIII) or as a result of the warp in \ngc 4013,
which is perpendicular to our line of sight (Bottema 1996) and therefore would not give
rise to the symmetrical distribution of clouds seen in our images.

There is a strong preference for the highest-\z\ absorbing structures
($\z\ \ge 2$ kpc) in \ngc 4013 to appear bent toward the northeast,
especially on the eastern side of the galaxy (see bottom panel in
Figure \ref{f-ngc4013v}). Such an arrangement could occur if there is
a rotational lag that increases with height (e.g. Heald \etal\
2007). This is consistent with the sense of rotation of \ngc 4013,
with the southwest side receding from us (Bottema 1996). \ngc 4013 is
a member of the spiral-rich Ursa Major group of galaxies.  Ursa Major
is considered a loose group with no central concentration and no sign
of strong \HI\ deficiencies, suggesting this group lacks a strong
intracluster medium (Angiras \etal\ 2007). As a member of the Ursa
Major group, \ngc 4013 is relatively isolated.  Thus the bending of
filaments is not likely the result of movement through an intracluster
medium in this case (Mart{\'{\i}}nez-Delgado \etal\ 2009). We find no
evidence for obvious supershell structures in \ngc 4013, although
identifying such structures within the complex distribution of clouds
might be difficult.

We see a set of very large-scale complexes of dust above and below the bulge of the galaxy
in Figure \ref{f-ngc4013v}. These large dust complexes, traced out to $\z\ \sim 1.8$ kpc
on either side of the bulge, show significant small-scale structure within them, including
dense clouds similar to those cometary shaped clouds identified in \pIII. Figure
\ref{f-ngc4013-cometary} shows one of these complexes to the southeast of the bulge.  This
large complex ($\sim900 \ {\rm pc} \, \times200$ pc) extends from within $\sim500$ pc of
the plane, bending toward the east with increasing height above the plane.  It shows
several cometary-like substructures at heights $\z\ \sim 600 - 850$ pc. Figure
\ref{f-ngc4013-cometary}(b) provides a close-up image of this large complex, which has a
mean extinction value of $a_{V} = 0.04$ and total mass of order $\ga 10^{5} M_{\odot}$.
Figure \ref{f-ngc4013-cometary}(c) shows two of the the cometary-like clouds, clouds
$115832.2+435644.6$ and $115832.0+435644.1$ from Table \ref{table:dustproperties}. 
Contours are included to guide the eye (but are based on the unsharp masked image and are
not quantitatively useful).  Cloud $115832.2+435644.6$ has a head-tail type structure; the
head of the cloud has $a_{V} = 0.36$ and implied density $\nH \geq 7$ cm$^{-3}$.  The
association of these cometary structures with the larger complex suggests the complex is
interacting with an ambient medium as it falls toward the disk through a corona or
outflowing wind from the disk.  

All of the head-tail clouds we see in \ngc 4013 seem to be associated with larger
complexes, like that shown in Figure \ref{f-ngc4013-cometary}.  The head-tail clouds do
not seem to be common, although the clouds may still be interacting with an ambient
medium.  Such morphologies are reminiscent of those seen in some high-velocity clouds
(HVCs) passing through the diffuse halo of the Milky Way. Putman \etal\ (2011)
investigated the \HI\ $21$ cm emission line morphologies of isolated HVCs, finding only
$\sim 35$\% of them show head-tail morphology, suggesting such morphologies are short
lived.  If the extraplanar clouds in our images are analogous to the HVCs, those with
cometary morphologies may be moving at larger velocities and interacting more strongly
with the thick disk ISM.  While the HVCs of the Milky Way do not appear to contain dust
with the exception of the Magellanic Stream (Lu \etal\ 1998, Sembach \etal\ 2001, Richter
\etal\ 2001, Fox \etal\ 2010), and their origins therefore are different from the thick
disk dusty clouds studied here, the physics that dictates the head-tail morphology is
likely quite similar. That is, the magnetohydrodynamical processes that shape those
cometary structures likely do not strongly depend on the dust content of the clouds.

\begin{figure}[!h]
\begin{center}
  \includegraphics*[scale=0.45]{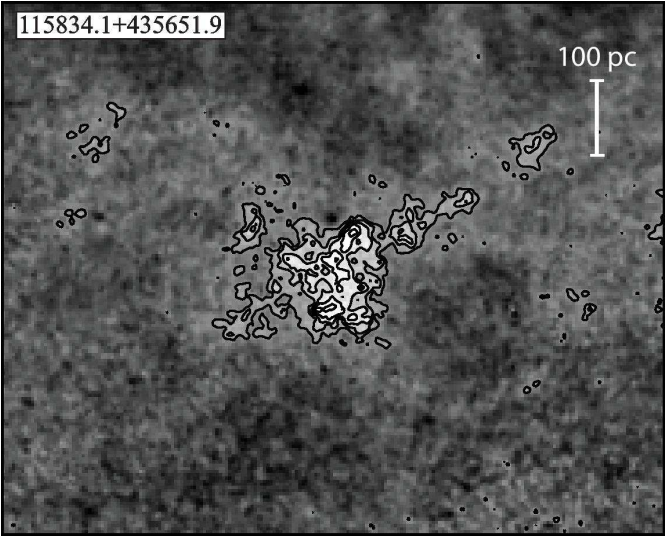}
  \caption{A close-up view of cloud $115834.4+435651.9$ in \ngc 4013. The orientation of
  this image is as in Figure \ref{f-ngc4013v}. This large cloud is located $\z\ \sim 930$
  pc from the midplane. The contours show $a_{V} = 0.2$ to $0.6$ in steps of $0.1$ mag. 
  As a single cloud we estimate the mass to be $M\ga$$7\times\e{4}$ $M_\odot$, although
  this depends on the definition of the cloud boundaries. This cloud shows prominent
  substructure. Two particularly dense dense subclouds at the top and bottom of the cloud
  have $a_{V}$ $\geq 0.5$ mag, implying column densities of $\Nh \geq 10^{21}$ cm$^{-2}$
  and particle densities of $\nH \geq 7$ cm$^{-3}$.} 
  \label{f-ngc4013-clump}
\end{center}
\end{figure}

Our \hst\ images reveal a vast array of small-scale absorbing features with dimensions
approaching our resolution limits.  The smallest-scale dust structures are smaller than a
few $\times\ 10$ pc in size. We have measured the sizes and opacities of a number of these
very small structures and estimate they have $a_{V} \sim 0.2 - 0.4$ mag and $\nH \geq 8$
cm$^{-3}$. Indeed, each complex exhibits significant substructure. To demonstrate this we
show in Figure \ref{f-ngc4013-clump} an expanded view of \ngc 4013 cloud
$115834.4+435651.9$, located at $\z\ \sim 930$ pc from the midplane. 
Prominent
substructures are clearly seen. The contours in Figure \ref{f-ngc4013-clump} range from
$a_{V} \sim 0.2 - 0.6$.  For comparison our spatially-averaged value for this cloud is
$a_{V} \sim 0.22$ (Table \ref{table:dustproperties}).  This demonstrates the impact of our
averaging technique when making our measurements. Two particular subclouds which are
similar in size and opacity, have $a_{V} \ga 0.5$ mag, implying column densities of $\Nh
\geq 10^{21}$ cm$^{-2}$ and particle densities of $\nH > 7$ cm$^{-3}$. These are lower
limits for the column densities and imply that these dust structures have properties
similar to Milky Way diffuse clouds, yet are located $\sim 1$ kpc away from the midplane.

While the dust in highly-structured clouds is most readily apparent in
our images, one can find very low-extinction, more diffuse material in
the thick disk in our images (particularly the WIYN and LBT images,
which have greater sensitivity to low surface brightness
fluctuations).  The lowest extinction clouds visible in our images of
\ngc 4013 have $a_V \sim 0.03$ mag.  If we couple that with a rough
measure of the area occupied by these low column density clouds (with
very low contrast), the total mass implied in this galaxy is of order
$\sim10^8$ \msun.  This is likely similar to the total mass in the
population of denser, more readily apparent complexes discussed above
(see HS00).

\subsubsection{\ngc 4302}
The extraplanar dust-bearing clouds in \ngc 4302 are filamentary with some large dust
complexes also notable in the thick disk in both the \hst\ and LBT images in Figure
\ref{f-ngc4302v}. The LBT unsharp masked display (bottom panel of Figure \ref{f-ngc4302v})
shows extensive extraplanar dust-bearing clouds along the entire length of the plane
detectable to $\z\ \sim 2.5$ kpc. While large complexes of extraplanar dust are seen on
both sides of the planes, the intensity of starlight as a function of \z\ shows a clear
asymmetry across the plane in the bulge region in \ngc 4302. The apparent deficit of
starlight at $\z\ \sim 1$ kpc on the west side of the plane corresponds to a large diffuse
complex of dust on this side of the disk. This diffuse complex is not easily seen in the
\hst\ unsharp masked images given its size and relatively little substructure, but is
visible in the unsharp masked LBT image (bottom panel of Figure \ref{f-ngc4302v}). This
extinction complex on the western side of the bulge area may be the cause for a measured
asymmetry in \Ha\ emission image as well (see \S\ref{sec:dig}).

A common characteristic in the majority of extraplanar clouds in \ngc 4302 is the apparent
bending of the large complexes toward the north at large \z -height, out to $\z\ \sim 2.3$
kpc.  This may be connected with the motion of \ngc 4302 through the Virgo Cluster, as
discussed by Chung \etal\ (2007).  These authors postulate the extensive tail of the H I
disk extending along the northern edge of \ngc 4302 is due to the galaxy's motion toward
the south as it falls into the Virgo Cluster.  Our images may be showing a physical
manifestation of this movement through the intracluster medium in the shaping of the
largest \z\ dust features. We note that \ngc 4298, projected close on the sky to \ngc
4302, is also located in the Virgo Cluster.  \ngc 4302 shows no sign of a strong
interaction with \ngc 4298.   

The ratio of thick-disk to thin-disk stellar scale heights is smaller in \ngc 4302 than
many other edge-on galaxies (as it is in the \Ha; Rand 1996). Combined with the
line-of-sight overlap of structures, this makes it difficult to cleanly identify large
scale extraplanar dust structures. Our images do not provide strong evidence for any
obvious supershells or galactic chimney features extending from the midplane of \ngc 4302.
Figure \ref {f-ngc4302_clouds} shows the unsharp masked \hst\ images of this galaxy with
the clouds from Table \ref {table:dustproperties} marked. As in \ngc 4013, we see
small-scale structures in the dusty clouds in \ngc 4302 to the resolution limit of our
\hst\ images ($\sim 10-20$ pc). The implied densities of the small-scale clouds range from
$\nH \geq 2$ to $14$ cm$^{-3}$ for the narrowest, darkest thick disk clouds.  

Similar to the case of \ngc 4013, there are diffuse, lower column
density structures seen over a significant amount of the thick disk
region of \ngc 4302.  These tend to have minimum extinctions of $a_V
\sim 0.02$ mag, implying a total thick disk mass in this more diffuse
matter of $\sim10^8$ \msun.

\begin{figure*}[!ht]
\begin{center}
  \includegraphics*[scale=0.9]{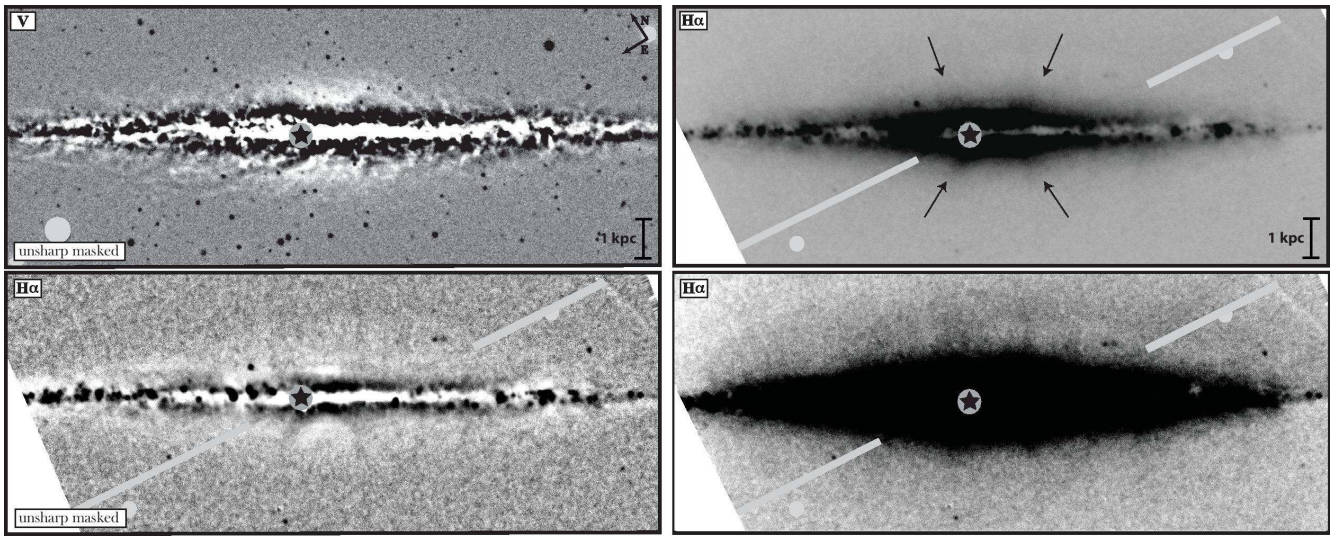}
  \caption{ Comparison of WIYN V-band and continuum-subtracted \Ha\ emission-line images
  of \ngc\ 4013.  The left panels shows the unsharp masked version of the WIYN V-band
  image (top) and the \Ha\ emission (bottom).  The right panels display the \Ha\ emission
  and have been stretch to emphasize the brightest (top) and faintest (bottom) \Ha\
  emission. Grayed areas mask artifacts contaminating the \Ha\ emission created by
  foreground stars, and some CCD artifacts. The prominent dust lane is clearly
  extinguishing \Ha\ emission within the plane of the galaxy. When comparing the
  extraplanar filamentary-like structures in the V-band image to the \Ha\ emission, there
  is little physical correlation between the features.  The unsharp-masked \Ha\ image, in
  particular, demonstrates that the few filaments found in the DIG are faint relative to
  the smooth DIG. The four faint filaments making up the large ``H''-shaped structure
  emanating from both sides of the bulge region are indicated by arrows in top-right
  panel. The filaments originate $R \sim 200-300$ pc from the center of the galaxy and can
  be traced in our images to at least $\z\ \sim 2.2$ kpc.} 
  \label{f-ngc4013ha}
\end{center}
\end{figure*}

\vspace{10pt}
\section{Ionized Gas in the Thick Disks of \ngc 4013 and \ngc 4302}
\label{sec:dig}

\begin{figure*}[!ht]  
 \begin{center}
 \includegraphics*[scale=0.9]{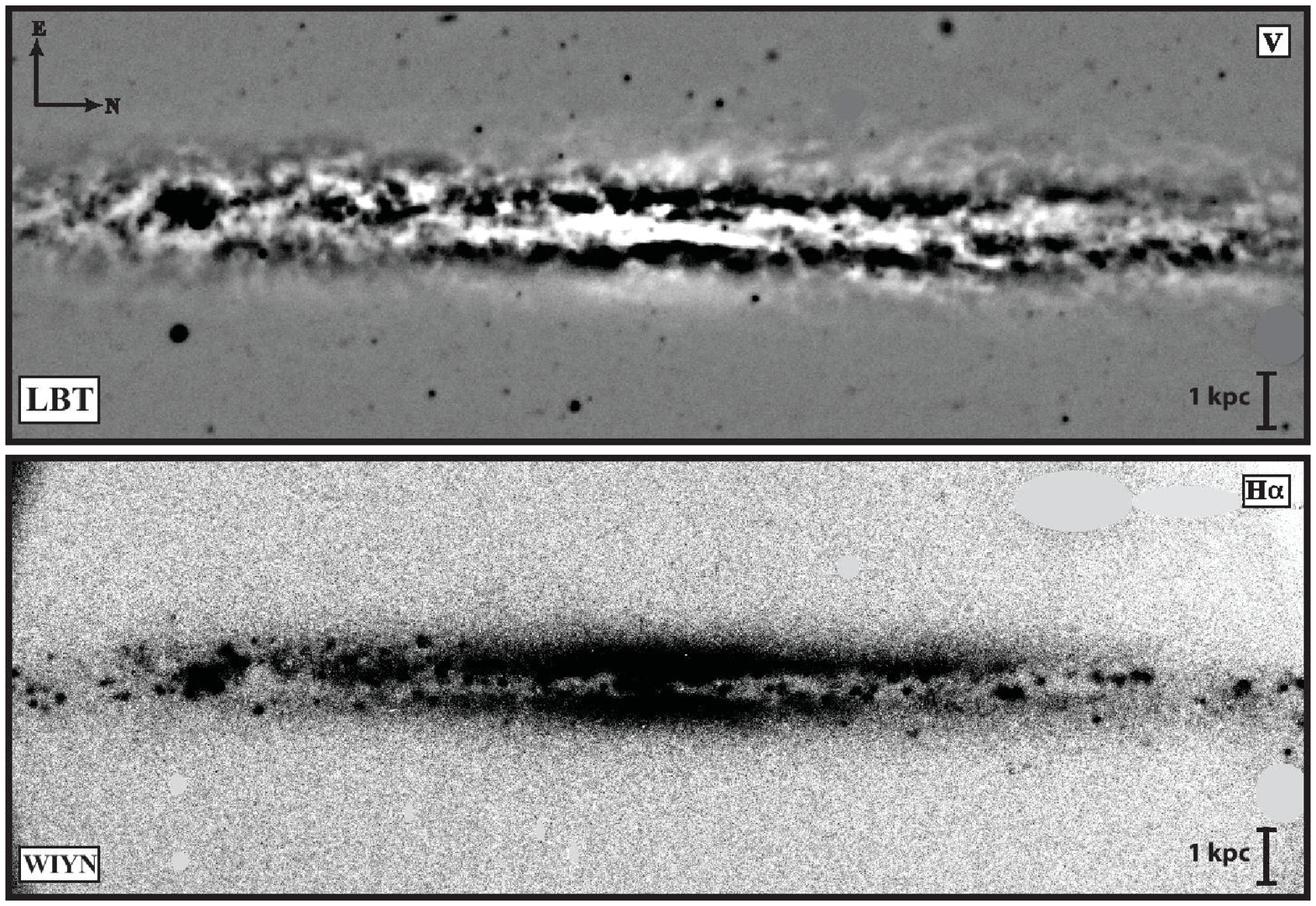}
  \caption{ LBT unsharp masked V-band image (top) and WIYN continuum-subtracted \Ha\ image
  of \ngc 4302 (bottom). Grayed areas mask artifacts due to saturated stars.  The
  extraplanar \Ha\ is weak in intensity except in the inner regions of the galaxy.
  Extraplanar absorbing dust structures seen in the V-band image are not reflected in the
  \Ha\ emission. There is little evidence for a direct physical connection between the
  extraplanar dust, which shows structure to very small scales, and the extraplanar DIG,
  which is very smooth.} 
  \label{f-ngc4302ha}
\end{center}
\end{figure*}

Extraplanar diffuse ionized gas (DIG) has been previously studied in both \ngc 4013 (Rand
1996, Miller \& Veilleux 2003) and \ngc 4302 (Pildis \etal\
1994, Rand 1996, Rossa \& Dettmar 2000, Collins \& Rand 2001, Heald
\etal\ 2007).  We have revisited these galaxies in an effort to achieve higher resolution,
greater depth, and/or more consistent continuum subtraction a previous works. Our
overall goal is to compare the \Ha\ and dust morphologies to study their connection in these
galaxies, which requires high resolution imagery. We detect thick disk \Ha\ emission, with
$0.8\arcsec$ resolution, in both galaxies.

\subsubsection{\ngc 4013}
Extraplanar DIG at large \z -heights can be seen over nearly the whole of the star-forming
disk (to R $\sim 9$ kpc) in \ngc 4013 (see the bottom-right panel of Figure
\ref{f-ngc4013ha}, which emphasizes the faintest \Ha\ emission). The extraplanar \Ha\
emission in \ngc 4013 is considerably fainter and shows much less prominent filamentary
emission than in \ngc 891 (Rand 1996).  Rand (1996) gives \Ha\ emission scale heights of
$h_z$ $\sim 0.7-0.9$ kpc (for the eastern side); however, he also notes significant
continuum subtraction difficulties affect his images. Miller \& Veilleux (2003) fit a
scale height of $h_z$ $\sim 1.5$ kpc (for the "southeastern" side). Figure
\ref{f-ngc4013ha} shows our \Ha\ images of \ngc 4013 in four panels; the top-left panel is
an unsharp-masked WIYN V-band image to show the dust absorption at the same resolution as
the \Ha\ emission while the bottom-left panel is an unsharp-masked \Ha\ image to show the
small scale features (as before, the smoothing kernel has FWHM$\, \la 350$). Both right
panels in Figure \ref{f-ngc4013ha} are \Ha\ emission, stretched to emphasize the brightest
emission (top-right panel) and the faintest emission (bottom-right panel). Our images are
consistent with the description of the DIG morphology for \ngc 4013 given by Rand (1996):
diffuse with few filaments.  Over the majority of the disk, the DIG morphology is smooth
and lacks bright, narrow filamentary structures like those seen in the \Ha\ emission of
\ngc 891 and \ngc 5775 (Rand 1996, \pIII). What filaments exist in Figure
\ref{f-ngc4013ha} are relatively subtle, the brightest of which have peak intensities of
$\sim 10-20 \%$ of the local smooth DIG. Even the brightest filaments in \ngc 4013 are
comparatively fainter than those in \ngc 891, which have $30 \% -100 \%$ of the brightness
of the local diffuse DIG emission (\pIII).  Given the faintness of the smooth DIG layer in
\ngc 4013, the filaments have a much lower absolute intensity as well. 

One of the most notable morphological features of the DIG in \ngc 4013 is the large
``H''-shaped structure emanating from both sides of the bulge region (see Rand 1996). This
``H'' structure consists of four large-scale filaments present at similar positions on
either side of the bulge, at projected radial distances $R \sim 200-300$ pc and visible in
our images to $\z\ \sim 2.2$ kpc. Rand (1996) argued this feature was a relic nuclear
(bipolar) outflow from an earlier nuclear starburstor AGN.  \garciaburillo\ \etal\ (1999)
mapped the ``H'' structure in CO emission (see their Figure 3) and argued for a close
association between the extraplanar DIG filaments and the extraplanar CO emission. 
Schinnerer \etal\ (2004) also mapped CO emission from the plane of \ngc 4013, connecting
the ``H'' structure in a high-density CO ring in the disk.  These works both invoke a superwind
outflow as the origin of this structure, in agreement with Rand's conclusion.

\subsubsection{\ngc 4302}

Figure \ref{f-ngc4302ha} shows the continuum-subtracted \Ha\ emission in \ngc 4302. The
images of this galaxy are not as deep as for \ngc 4013, but they have better resolution
than most previous \Ha\ images of this galaxy (Pildis 1994, Rand 1996, Rossa \& Dettmar
2000, 2003a, 2003b). \ngc 4302 has one of the faintest detectable DIG layers (Rand 1996,
Rossa \& Dettmar 2000) Collins \& Rand (2001) measure the emission scale height $h_z \sim
0.55 - 0.7$ kpc. Rand (1996) characterizes the extraplanar DIG morphology of \ngc 4302 as
``faint and diffuse"; our images are consistent with both of those characterizations. In
our images the thick disk DIG is smoothly distributed within the first $\z\ \sim 2$ kpc of
the plane, but only within R $\sim 7$ kpc and $8.5$ kpc radially from the bulge in
northern and southern halves of the galaxy, respectively. The \Ha\ emission abruptly
decreases to undetectable levels at larger radii. We detect slightly brighter extraplanar
DIG on the eastern side compared to the western side of the galaxy, most notably around
the bulge region. This asymmetry is also seen in continuum star light and is due to
enhanced extinction from large extraplanar dust complexes on the western side of the plane
at $\z\ \sim 1$ kpc (discussed in \S 3.3).  We find no bright plumes or filamentary
structures. There are some very weak DIG filaments, the brightest of which measure only
$\sim 20-25 \%$ above the local (very faint) diffuse background. Thus, the DIG in \ngc
4302 shows little to no structure or filaments, and what substructure is detected seems to
be the result of dust extinction affects.

Dettmar (1995) and Heald \etal\ (2007) found that the presence of \Ha\ emission in \ngc
4302 was weak and diffuse, especially in comparison to \ngc 891, although both galaxies
have prominent extraplanar dust.  Our \Ha\ images are the highest resolution available for
both \ngc 4013 and \ngc 4302 and show both have smoothly distributed DIG layers with much
less prominent filaments than the DIG of \ngc 891 and even less than \ngc 5775 (Rand 1996,
\pIII).  Both \ngc 4013 and \ngc 4302 also have comparably weaker \Ha\ emission than \ngc
891 and \ngc 5775, which are known for their bright \Ha\ emission and have larger DIG
emission scale heights (Collins \& Rand 2001). Thus the galaxies can be ranked in order of
decreasing DIG scale height: \ngc 5775, \ngc 891, \ngc 4013, \ngc 4302. This is the same
order one would have if ranking by decreasing \Ha\ intensity and decreasing prominence of
DIG filaments.
\section{The Relationship Between Extraplanar Ionized Gas and Dust}
\label{sec:comparison}
	
	\ngc 4013 and \ngc 4302 show both dust and DIG in the thick disk. Is there any
	relationship between the physical structures probed by these tracers of the thick disk
	ISM? \ngc 891 is the only galaxy for which a detailed comparison between the
	dust-bearing cloud structures and the DIG in the thick disk has been made at high
	resolution, although several works have noted the distinct morphologies of thick disk
	dust and DIG in \ngc 891 (Dettmar 1990, \pI, \pIII, Rossa \& Dettmar 2003a, 2003b,
	Rossa \etal\ 2004, \ths). Howk \& Savage (\pIII) argued the dusty clouds in the thick
	disk of \ngc 891 trace a dense medium, in part because of these differing
	morphologies. They also estimated typical densities for the extraplanar dust-bearing
	clouds of $\nH \geq  2 -10 $ cm$^{-3}$, significantly higher than DIG structures at
	any distance from the midplane. Furthermore, \pIII\ found very few cases where
	structures traced in \Ha\ emission and dust absorption could be even loosely
	associated. They found no direct physical connection between the extraplanar
	dust-bearing structures and the DIG seen in \Ha\ emission. Rossa \etal\ (2004) reached
	the same conclusion after comparing the thick disk dust extinction and \Ha\ emission
	in high resolution \hst\ imaging of \ngc 891. Keppel \etal\ (1991) also noted the
	emission measure from clouds as dense as those probed by the extraplanar dust should
	be much higher than that observed from the DIG in \ngc 891, suggesting the dusty
	clouds trace a different medium.

Our observations allow us to explore the relationship between extraplanar dust extinction
and \Ha\ emission at high resolution in two additional galaxies. Extraplanar dust
filaments are very prominent in the thick disk of both galaxies, whereas the extraplanar
DIG layer lacks prominent filaments or structures in both.  Figures \ref{f-ngc4013ha} and
\ref{f-ngc4302ha} show the \Ha\ emission images on the same physical scales as the
broadband ground based images, allowing for a direct comparison of the dense thick disk
dust cloud structures and the \Ha\ emitting DIG structures.
		
	In \ngc 4013 we see filamentary dust structures in the thick disk in our broadband
	\hst\ and WIYN images along the entire radial length of the galaxy, with extensive and
	highly-structured dust filaments seen near the bulge region.  Our images show
	absorbing dust structures out to heights $\z\ \sim 2.0$ kpc on both sides of the plane
	of \ngc 4013. On the largest scales, the extraplanar dust and DIG show distinct
	distributions. The DIG is relatively smooth with weak and few filaments.  The DIG
	filaments that do exist are not reflected in the dust. The large extraplanar dust
	complexes themselves are not directly identifiable in the \Ha\ emission. The lack of a
	morphological connection between extraplanar dust clouds and the DIG is consistent
	between the largest and smallest scales in \ngc 4013.

	In \ngc 4302 we come to the same conclusion: the morphologies of the absorbing dust
	structures and \Ha\ emitting structures show no obvious physical relationship. In \ngc
	4302 our broadband images show filamentary dust-bearing clouds extending to \z\ $\sim
	2.5$ kpc along the length of the galaxy. The thick disk DIG layer has very few
	filamentary structures and is smoothly distributed along the inner R(north) $\sim 7$
	kpc and R(south) $\sim 8.5$ kpc; at larger R the extraplanar \Ha\ emission quickly
	disappears.  This drop-off in the DIG emission contrasts with the behavior of the
	extraplanar dust filaments, which are seen extending to large \z -heights to nearly
	the edge of the optical disk.  The \Ha\ morphology we see in \ngc 4302 is a result of
	large dust complexes extinguishing emission, yet no physical DIG structures are traced
	by any dust filaments seen in our \hst\ and LBT images. Overall, the dust clouds are
	structured on very small scales and strongly filamentary, while the extraplanar
	ionized gas is primarily diffuse and smooth.

When comparing the morphologies of the dust and DIG in the thick disk, neither \ngc 4013
nor \ngc 4302 show evidence for a direct physical relationship between the thick disk dust
seen in the broadband images and the extraplanar ionized gas seen in \Ha\ emission. \ngc
4302 shows the strongest contrast in the physical appearance of the dust and DIG as it has
one of the weakest extraplanar DIG layers yet has a very dusty thick disk. In a sense, we
expect very little connection between the dust structures and DIG given the importance of
the smooth components of the DIG in these two galaxies (Rand 1996, Rossa \& Dettmar 2000,
Heald \etal\ 2007). The smaller-scale structures in the dust-bearing clouds, their larger
column densities, and even the differences in their large-scale distribution emphasize
that the dust traces a phase of the thick disk ISM distinct from the DIG.

\ngc 4013 and \ngc 4302 are two edge-on galaxies that display a lack of a direct physical
correlation between thick disk dust and ionized gas. This physical difference between the
morphology of the extraplanar DIG and dust is not unique to \ngc 891. At the same time,
there is a correlation between the presence of extraplanar dust and DIG disks of galaxies.
As first noted by \pII\ in a small sample of galaxies and subsequently in the much larger
survey of Rossa \& Dettmar (2003a, 2003b), when DIG is detected in a galaxy, extraplanar
dust is also visible $\sim $ 90\% of the time. Thus, although there is no evidence for a
morphological relationship between the dust and DIG, there is a strong correlation between
the presence of extraplanar dust and DIG in galaxies, a result reflected in \ngc 4013 and
\ngc 4302.

\section{The Multiphase Thick Disks of Spiral Galaxies}
\label{sec:discussion} We have presented deep broadband and \Ha\ images of the edge-on
spiral galaxies \ngc 4013 and \ngc 4302.  Our images reveal significant numbers of dusty
extraplanar clouds via the extinction they produce against the background thick disk
and bulge starlight in both galaxies.  These galaxies were chosen in part because the
morphology, brightness, and scale height of the DIG emission in these galaxies was
significantly different than the other galaxy for which a detailed comparison of the
extraplanar dust and DIG is available, \ngc 891 (\pIII, Rossa \etal\ 2004).  Thus, our
program is designed to understand whether the changes in the morphology and properties of
the DIG are reflected in any way in the extraplanar dust.  While various studies have
argued that the thick disk dust absorption is tracing a different medium or phase than the
DIG (Keppel 1991, \pII, \pIII, Rossa et al. 2004) with distinct morphologies, we are
testing whether this dense medium responds at all to the processes that shape the DIG
properties.

We find no evidence that the morphologies and properties (quantity, sizes, masses, etc.)
of the extraplanar dust-bearing clouds are subject to the strong changes like those seen
in the DIG morphologies. Table \ref{table:comparison} presents $L_{FIR}$, star formation
rate (SFR), and star formation surface density ($\dot{\Sigma}_{*}$) for \ngc 4013 and \ngc
4302 in comparison to previously studied edge-on galaxies \ngc 5775, \ngc 891, and \ngc
4217. The star formation surface density (SFR per unit area) $\dot{\Sigma}_{*} \equiv$
SFR/$A_{25}$, where $A_{25} = \pi D^2_{25}$/$4$. Ordering these galaxies by SFR (derived
from the $L_{FIR}$) Table \ref{table:comparison} shows a connection between the morphology
of the extraplanar DIG and the SFR. Thus, as the importance (prominence, brightness, etc.)
of the DIG filaments and scale height decrease along the sequence of \ngc 5775, \ngc 891,
\ngc 4217, \ngc 4013, and \ngc 4302, the morphological and physical properties of the
observable dust seem to vary little.  This is in part a selection effect, since the
detection of the extraplanar dust in our images relies on the small-scale structure of the
absorbing structures.  The thick disk dust is only visible because it is more opaque than
its surroundings, and this likely implies the clouds in our images have significantly
higher particle densities than their surroundings. However, this also emphasizes our
point: the properties of the dust are disconnected from that of the DIG in the thick disk.

In fact, one can see morphological and structural changes in the DIG that are not mirrored
in the extraplanar dust within individual galaxies. In \ngc 891 the southwest half of the
disk has a smaller DIG scale height, with many fewer and less prominent filaments than the
northeastern half of the disk (\pIII).  However, there is relatively little change in the
properties and incidence of dust between these regions, save that there are a very few DIG
filaments in the northeastern half that are physically related to extraplanar dust
structures.  (Such an arrangement is only seen in this portion of \ngc 891 and in some
portions of the strongly-star forming galaxy \ngc 4631; see Wang et al. 2001.) Similarly,
the present observations show the DIG layer in \ngc 4302 cuts off sharply at a projected
radial distance from the nucleus of R $\sim 7$ -- $8.5$ kpc, while the extraplanar dust
can be seen well beyond these projected radii.  Additionally, the thick disk near the
bulge region of \ngc 4013 shows highly structured filaments whereas the extraplanar DIG
morphology is smooth even in the brightest \Ha\ emission.

Apparently the factors that shape the morphology and quantity of the extraplanar DIG do
not affect the dusty extraplanar clouds seen in our images.  While there is a strong
correlation between the existence of extraplanar DIG and extraplanar dust within a galaxy
(\pII, Rossa \& Dettmar 2003b), the ionized gas traced by \Ha\ emission and dense clouds
traced in optical continuum absorption have very little to do with one another.  What
drives this distinct behavior of the DIG and dense, dusty clouds?  One possibility is that
the SFR or star formation surface density (SFR per unit area) plays a role in shaping the
morphology of the DIG layer. Rossa \& Dettmar (2003a) found DIG morphology was connected
to star formation activity in a galaxy, with the transition from higher to lower
$\dot{\Sigma}_{*}$ or SFR being associated with a progression from DIG with prominent
filamentary structures through fainter, more diffuse DIG. The sequence then of \ngc 5775,
\ngc 891, \ngc 4217, \ngc 4013, to \ngc 4302 is one of decreasing SFR as well as
decreasing DIG scale height and \Ha\ filament prominence.  While this is a limited sample,
we also see changes in star formation affect the DIG morphology on smaller scales. Rossa
\& Dettmar (2003a, 2003b) note that star formation on the local and global scales were
factors in the presence of extraplanar DIG in their survey.  For example, the southwestern
half of \ngc 891 has a smaller DIG scale height and fewer prominent filaments than the
northeastern half (\pIII) while also having a lower SFR as evidenced by radio continuum
and far infrared measurements (Dahlem \etal\ 1994, Wainscoat \etal\ 1987). The edges of
\ngc 4302 beyond the end of the DIG layer are also representative of this trend.  

By what mechanism could the star formation or its surface density affect the morphology of
the DIG in this way?  The answer is unclear.  It could be that the filaments seen in the
DIG are tracing ionized fragments of supershells whose presence and brightness require
very recent star formation in clusters to maintain the ionization, while the smoother DIG
component is ionized by longer-lived field OB stars (e.g., Hoopes \& Walterbos 2000).
However, Rand (1998) found that the bright filaments in \ngc 891 did not show unusual line
ratios (e.g., [\ion{N}{2}]/\halpha, [\ion{S}{2}]/\halpha) for the DIG.  In fact, the
filaments seemed to show the same overall behavior as the more diffuse DIG, with lower
forbidden line to \halpha\ ratios associated with brighter emission. Thus, if the
filaments are associated with recent superbubbles, they still share common ionization and
excitation characteristics with the general DIG.  At this point this assignment of cause
is largely speculation. It seems likely that SFR and/or $\dot{\Sigma}_{*}$ likely play a
significant role in not only determining whether a galaxy has significant DIG emission
(Rossa \& Dettmar 2003a, 2003b, Rand 1996), but also in shaping the DIG structure (Rossa
\& Dettmar 2003b).

As with the extraplanar DIG, a minimum star formation rate or $\dot{\Sigma}_{*}$ seems to
be required before extraplanar dust is observable (Rossa \& Dettmar 2003a, 2003b), but
above that threshold there is little discernable effect on the morphology (Table
\ref{table:comparison}).  Furthermore, there is no evidence that the positional changes in
$\dot{\Sigma}_{*}$ that modify the DIG morphology affect the dense extraplanar clouds. In
comparing the DIG morphology sequence outlined in Table \ref{table:comparison} to the dust
morphologies, the DIG changes in connection with the SFR, while the extraplanar dust is
both present and consistently similar between all of the galaxies. That the dust clouds
have little associated \Ha\ emission likely means they are not directly illuminated by
ionizing radiation (Keppel et al. 1991), since they are optically thick to such radiation.
 Perhaps they have a different evolutionary timescale for production than the DIG (e.g.,
Alton et al. 2000), and the stars that would have ionized the clouds have already died
off. It may also be that they have a distinct radial distribution compared with the DIG. 
We see only the foreground dust-bearing clouds with the direct imaging techniques, but
perhaps there are few such clouds at the smaller radial distances at which the DIG
dominates.  It is unlikely we could show this without velocity information for the
extraplanar dust clouds, which is not likely unless they can be measured through CO
emission (e.g., \garciaburillo\ et al. 1999).  

Discussions of the origin of the matter making up the dense clouds seen in our images and
the phase structure of these clouds have been presented elsewhere (\pI, \pII, \pIII, Howk
2005).  Here we expand on those discussions briefly to reflect new thinking (both our own
and in the literature).  The interstellar thick disk clouds seen via their dust absorption
within $z \sim2$ kpc of the midplanes of a number of spiral galaxies are almost certainly
dominated by material that has been expelled from the thin disk via stellar feedback,
perhaps both mechanical and radiative.  In this case, the signpost used to trace the
matter -- the dust -- identifies the clouds as arising from metal-rich material.  This has
two important implications.  First, the great quantities of dust in these clouds implies
that they are dominated by metal-rich thin disk matter that has been expelled rather than
by infalling (metal poor) matter as invoked to explain the change in DIG rotation curves
with \z\ (e.g., Fraternali \& Binney 2006, Heald et al. 2007).

Second, the presence of dust-rich matter in the thick disk implies that the processes that
lift the dust from the thin disk are not so violent that they destroy a large fraction of
the dust. If a significant amount of dust is destroyed, the masses calculated for our
observed clouds scale upwards.  If only 10\% of the dust survived the trip to the thick
disk the total mass in these clouds would approach the total gas mass of the ISM in these
galaxies (\pIII), which is unrealistic.  The existence of dust in the thick disk fueled by
feedback-driven expulsion is clearly not a new discovery. It does not cause difficulties
with existing models for galactic fountain-type phenomena.  The walls of a supershell may
extend to quite large distances from the plane, but much of the mass was incorporated into
those walls while the shell expanded relatively slowly (e.g., Weaver et al. 1977)
destroying little of the associated dust.  Furthermore, it is possible that radiation
pressure plays a role in this process (Murray et al. 2011, Franco et al. 1991), which
would leave all of the dust intact. A completely independent method of probing the
existence of dust surviving far out of the plane is through the infrared wavelengths. \ngc
891 has detectable mid-infrared radiation out to $z \sim5$ kpc, which is dominated by
emission from diffuse interstellar dust (Burgdorf, Ashby, \& Williams 2007).  Thus,
extraplanar dust is present many kpc further from the star-forming disk than our images
reveal. Its existence to $z \sim5$ kpc (and even beyond; see M\'{e}nard \etal\ 2010)
implies the dust is not all destroyed as it is lifted from the thin disk. 

The thick disk clouds traced by the extraplanar dust absorption in our
optical images are physically distinct from the DIG.  They trace material with densities
larger than that of the DIG by an order of magnitude or two.  The large gas densities
($n_{\rm H} \ga 1-25$ \percc ; \pIII, Howk 2005, this work) and high gas column densities
($\Nh \ga 10^{20}$ to $\ga 10^{21}$) traced by the extraplanar dust are largely consistent
with expectations for gas associated with a warm neutral or cold neutral medium.  Both are
likely present in these clouds, given the level of substructure and evidence for core-halo
type structure in some clouds (see also \pIII). Although in principle one might be able to
use the distribution of apparent extinction in isolated regions of these galaxies to
disentangle the relative contributions of these phases, the results are ambiguous given
the fitting uncertainties in the background light distribution and the presence of clouds
deeper in the galaxy with a larger fraction of unabsorbed starlight.

\pIII\ and \pII\ have argued that the extraplanar dust may to a large degree trace a thick
disk CNM based on assessments of the physical properties and a comparison of the cloud
masses with Galactic molecular clouds (Blitz 1990, 1991).  There is complementary evidence
in our work for the presence of a thick disk CNM in the form of the extraplanar CO
emission detected in \ngc 4013 (\garciaburillo\ et al. 1999) and the probable presence of
thick disk \HII\ regions tracing recent star formation (\pIII, Rossa \& Dettmar 2003a).
Candidate thick disk \HII\ regions can be seen in both \ngc 4013 and \ngc 4302 (point-like
knots of \Ha\ emission visible in Figure \ref{f-ngc4013ha} and \ref{f-ngc4302ha}).  In
particular, Figure \ref{f-ngc4013ha} shows a prominent candidate to the north of the bulge
of \ngc 4013 (spectroscopy of this nebula is focus of K. Rueff \etal, in preparation).
Potential areas of star formation suggest a CNM as a component of the thick disk ISM in
these galaxies. While the masses of the clouds reported in Table
\ref{table:dustproperties} are somewhat lower than those studied in earlier works, this
difference is in part due to the conscious choice to study smaller clouds in order to
emphasize the small-scale structure revealed by our high resolution images.  The
properties of the extraplanar dust-bearing structures in our images are consistent with a
significant CNM present in the thick disks of these spiral galaxies. However, the fact
that the clouds vastly outnumber the \HII\ regions so far identified suggests that the
fraction of the former that trace molecular clouds able to support star formation (as
opposed to ``diffuse clouds'' in the Galactic parlance) may be small.

While the thick disk gas traced by these extraplanar dust structures largely originates in
the thin disk, the production or maintenance of a thick disk CNM can come about in several
ways.  \pIII\ and \pII\ discussed the possibility that the CNM is consistent with standard
theories for multiphase thermal equilibrium of the ISM (e.g., McKee \& Ostriker 1977,
Wolfire et al. 1995a,b, Wolfire et al. 2003). Here we add that more dynamical origins may
be even more important. As feedback-driven outflows escape from the thin disks of
galaxies, they sweep up significant masses of material. Some of this material is
compressed at a rate insufficient to shock the gas to high temperatures and/or at a time
when the shell walls might radiatively cool.  In this case, the shell serves to compress
the gas, triggering the formation of a cold or even molecular medium (e.g., McCray \&
Kafatos 1987).  This is, of course, the origin of the self-propagating star formation
theories.
  
In a similar vein, recent work has highlighted the instabilities that can arise in
converging flows of neutral material. This can serve to compress and cool the gas (e.g.,
Heitsch et al. 2008), and the convergence of the walls of adjacent shells may be a trigger
for cloud formation (Ntormousi et al. 2011).  Our observations of extraplanar dust always
show numerous and very complicated structures within the first kpc of the galaxy
midplanes, with few readily-identifiable supershells. Some of this is undoubtedly due to
line of sight confusion.  However, given the difficulty of distinguishing physically
connected structures in our images, it is not a stretch to imagine some of the complexity
we see in the dust occurs as walls of adjacent supershells or other structures converge.
Converging flow-type scenarios provide a sustainable method for forming CNM clouds in the
thick disk and for reshaping the structures in a less-ordered fashion. The precise
mechanism of the formation of CNM clouds in the thick disk is little constrained by our
data.  However, we note that the associations of cool (Moss et al. 2012) or even molecular
gas (Dawson et al. 2011a,b) with high-latitude supershells in the Milky Way offer
supporting evidence that the ejection of the gas from the thin disk may also be in part
responsible for triggering the formation of a cool medium.  The global characteristics and
physical conditions for the material traced by the extraplanar dust in our images suggest
the presence of a thick disk CNM in addition to warm neutral and ionized gas. In future
works we will explore the thick disk star formation, perhaps from this CNM gas, implied by
extraplanar \HII\ regions.

\section{Summary}
\label{sec:summary}

We have presented \hst\ /WFPC2 and ground-based WIYN 3.5-m and LBT $2 \times 8.4$ m images
of the edge-on spiral galaxies \ngc 4013 and \ngc 4302.  These images were used to study
the distribution of extraplanar dust and ionized gas in these galaxies.  The major results
of our work are as follows.

1. Our new images reveal extensive distributions of extraplanar dust clouds in \ngc 4302
and \ngc 4013, with absorbing structures traced to heights \z\ $\sim 2$ kpc in both
galaxies.  The thick disk dust-bearing gas clouds are seen in large complexes 
as well as narrow filaments, with significant substructure seen to the limit 
of our \hst\ resolution ($< 10-20$ pc).

2.  We derive the properties of several extraplanar dust clouds in each galaxy (Table
\ref{table:dustproperties}).  We estimate hydrogen column densities for individual dust
clouds N(H) $\ga 2 \times 10^{20}$ cm$^{-2}$ and masses $\ga 10^{4}  M_{\odot}$, with the
larger complexes having quite a bit higher masses.  The implied particles densities range
from $\nH > 1.0-4.5$ cm$^{-3}$ for the larger clouds, with substructure clouds having even
larger densities, $\nH > 7$ cm$^{-3}$. The properties of these individual absorbing
structures lead us to conclude the dust clouds trace a dense phase of the thick disk ISM.

3.  Our high-resolution \Ha\ images show the extraplanar ionized gas layers of both
galaxies have a diffuse, smooth distribution, with very few filamentary structures.
Candidate thick disk \HII\ regions can be seen in both \ngc 4013 and \ngc 4302.  In
\ngc 4302, the \Ha\ emission abruptly cuts off at projected radii of R $\sim 7$ kpc and
$8.5$ kpc in the northern and southern halves of the galaxy, respectively.

4.  Comparing the morphologies of the extraplanar dust clouds to the extraplanar DIG
emission we find little evidence for a direct physical connection between the dust and the
DIG. The dust-bearing clouds are strongly filamentary  while the DIG layers lack
filaments and are smoothly distributed. The features and physical properties of the
extraplanar dust and DIG in \ngc 4013 and \ngc 4302 are distinct.

5.  The thick disk ISM extending $\sim 2$ kpc away from galactic disks is a multi-phase
system, including a dense, cold phase traced by dust extinction and perhaps formed through
thermal instabilities and/or in converging flows.  This thick disk in \ngc 4013 and \ngc
4302 is ultimately fueled by galactic fountain-like expulsion from the star-forming thin
disk that does not destroy the dust as it raises material to high-\z.

\acknowledgements
We thank an anonymous referee for helpful comments on the paper. KMR and JCH recognize
support from NASA through grant NNX10AE87G. Support for \hst\ program number 8242 was
provided by NASA through a grant from the Space Telescope Science Institute, which is
operated by the Association of Universities for Research in Astronomy, Incorporated, under
NASA contract NAS5-26555.

This research has made use of the NASA/IPAC Extragalactic Database (NED) which is operated
by the Jet Propulsion Laboratory, California Institute of Technology, under contract with
the National Aeronautics and Space Administration.  This research has made use of the
VizieR catalogue access tool, CDS, Strasbourg, France.

\pagebreak
\clearpage
\newpage

\begin{deluxetable}{lccccccccc}[!h]
\centering
\tablenum{1} 
\tablewidth{0pc}
\tablecolumns{10}
\tablecaption{General Characteristics of \ngc 4013 and \ngc 4302\tablenotemark{a}
\label{table:galaxies} }
\tablehead{\colhead{Name} & \colhead{RA} & \colhead{Dec.} & \colhead{$D_{25}$} & 
\colhead{V$_{rad}$} & \colhead{$W_{20}$} & \colhead{Dist.\tablenotemark{b}} &
\colhead{SFR\tablenotemark{c}} & \colhead{Type} \\
	\colhead{} & \colhead{[J2000]} & \colhead{[J2000]} & 
	\colhead{[arcmin]} & \colhead{[km s$^{-1}$]} & 
	\colhead{[km s$^{-1}$]} & \colhead{[Mpc]} & 
	\colhead{[$M_\odot $yr$^{-1}$]} & 	\colhead{} 	}
\startdata
NGC 4013  & 11 58 31 & +43 56 48 & 4.7  & 831  & 407 &  17.0  &  1.5 &  Sb  \\ 
NGC 4302  & 12 21 42 & +14 35 54 & 4.7  & 1149 & 377 & 16.8  & 0.9 &  Sc  \\
\enddata
\tablenotetext{a}{The properties presented here were taken from the NED database, except
	where noted.} 
\tablenotetext{b}{We adopt distances from the Extragalactic Distance Database (Tully
	\etal\ 2009).} 
\tablenotetext{c}{The star formation rate, SFR, were calculated following Kewley \etal\
(2002), using the $L_{FIR}$ and the Kennicutt (1998) SFR calibration (see SFR discussion
in \S \ref{sec:discussion}).}  
\end{deluxetable}

\newpage
\begin{deluxetable}{rrcc}[!h]
\centering
\tablenum{2} 
\tablewidth{0pc}
\tablecolumns{4}
\tablecaption{Log of HST/WFPC2 Observations\tablenotemark{a} \label{table:hstlog}}
\tablehead{
\colhead{Filter}  & \colhead{}  & \multicolumn{2}{c}{Total Exp. Time [seconds]} \\
\cline{3-4}
\colhead{} & \colhead{} & \colhead{\ngc 4013\tablenotemark{b}} & \colhead{\ngc 4302}} 
\startdata
 		 F450W ($B$-band)  & & 4000 &  4000  \\    
		 F555W ($V$-band)  & & 3000 &  2000  \\
		 F814W ($I$-band)  & & 3000 &  2000  \\ 
\enddata
\tablenotetext{a}{The angular and physical resolutions are $0\farcs1$ and $8$ pc, 
respectively.}
\tablenotetext{b}{The \ngc 4013 images have non-uniform exposure times per pixel. Two
separate visits with \hst\ had distinct pointing centers and orientations.  The central
and northern sections of the galaxy, respectively, have three and two times the exposure
time of the southern-most region.}
\end{deluxetable}

\begin{deluxetable}{cccc}[!h]
\centering
\tablenum{3} 
\tablewidth{0pc}
\tablecolumns{4}
\tablecaption{Log of WIYN Observations\tablenotemark{a} \label{table:wiynlog}}
\tablehead{
\colhead{Filter} &
\colhead{Total Exp. Time}   & \colhead{Seeing\tablenotemark{b}} &
\colhead{Resolution\tablenotemark{c}} \\
\colhead{} &\colhead{[seconds]} &
\colhead{[arcsec]} & 
\colhead{[pc]} }
\startdata
\multicolumn{4}{c}{NGC 4013} \\
\cline{1-4}
		  $B$ & 2250 & 0.6 & 49 \\
		  $V$ & 2900 & 0.9 & 74\\
		  $I$ & 1800 & 0.8 & 66\\ 
		  $r$ & 3600 & 0.8 & 66\\
         		 \Ha\tablenotemark{d} & 14400 & 0.8 & 66 \\
\cline{1-4}\\
\multicolumn{4}{c}{NGC 4302} \\
\cline{1-4}
		$r$ & 3600 & 1.0 & 82 \\
        		 \Ha\tablenotemark{d} & 9000 & 0.8 & 65 \\
\enddata
\tablenotetext{a}{BVI-band data were taken with the S2KB detector. \Ha\ and $r$-band data
were taken with the WTTM.}
\tablenotetext{b}{Seeing values empirically measured in the final images.}
\tablenotetext{c}{Resolution based on assumed distances from Table \ref{table:galaxies}.}
\tablenotetext{d}{WIYN W015 filter}
\end{deluxetable}

\begin{deluxetable}{cccc}[!h]
\centering
\tablenum{4} 
\tablewidth{0pc}
\tablecolumns{4}
\tablecaption{Log of LBT/LBC Observations of \ngc 4302 \label{table:lbtlog}}
\tablehead{
\colhead{Filter} &
\colhead{Total Exp. Time}   & \colhead{Seeing\tablenotemark{a}} &
\colhead{Resolution\tablenotemark{b}} \\
\colhead{} &\colhead{[seconds]} &
\colhead{[arcsec]} & 
\colhead{[pc]} }
\startdata
$U$ & 2000 & 1.2 & 98\\
$B$ & 1000 & 1.1 & 90\\
$V$ & 1000 & 1.0 & 81\\
$I$ & 2000 & 0.9 & 73\\
\enddata
\tablenotetext{a}{Seeing values empirically measured in the final images.}
\tablenotetext{b}{Resolution based on assumed distances from Table \ref{table:galaxies}.}
\end{deluxetable}

\begin{deluxetable}{ccccccccc}[!h]
\centering
\tabletypesize{\scriptsize}
\tablenum{5} 
\tablewidth{0pc}
\tablecolumns{7}
\tablecaption{Properties of Individual High-$z$ Dust Features 
	\label{table:dustproperties}}
\tablehead{
\colhead{Cloud\tablenotemark{a}} & 
\colhead{Dimensions} & \colhead{$|z|$\tablenotemark{b}} & 
\colhead{$a_{V}$\tablenotemark{c}} & \colhead{$N_{\rm H}$\tablenotemark{d}} & 
\colhead{$n_{\rm H}$\tablenotemark{e}} & \colhead{Mass\tablenotemark{f}}  \\
\colhead{[J2000]} &
\colhead{[pc$\times$pc]} & \colhead{[pc]} &
\colhead{[mag.]} &  \colhead{[cm$^{-2}$]} &
\colhead{[cm$^{-3}$]} & \colhead{[M$_\odot$]} }
\startdata
\multicolumn{7}{c}{NGC 4013} \\
\cline{1-7} 
115835.5$+43$5702.0 & 90 $\times$ 60 
		& 590 & 0.20 & $>$$4$$\times$$\e{20}$ & 2.1 & $>$$3$$\times$$\e{4}$ \\
115834.4$+43$5713.5 & 50 $\times$ 45 
		& 640 & 0.17 & $>$$3$$\times$$\e{20}$& 2.2 & $>$$1$$\times$$\e{4}$ \\
115833.8$+43$5713.8 & 70 $\times$ 70
		& 890 & 0.15 &$>$$3$$\times$$\e{20}$& 1.3 & $>$$2$$\times$$\e{4}$ \\
115834.4$+43$5651.9 & 170 $\times$ 150 
		& 930 & 0.22 & $>$$4$$\times$$\e{20}$& 0.9 &  $>$$7$$\times$$\e{4}$ \\
115832.9$+43$5706.8 & 40 $\times$ 30 
		& 690 & 0.22 &$>$$4$$\times$$\e{20}$& 4.5 &  $>$$5$$\times$$\e{3}$ \\
115833.5$+43$5648.8 & 90 $\times$ 55 
		& 850 & 0.12 &$>$$2$$\times$$\e{20}$& 1.4 & $>$$1$$\times$$\e{4}$ \\
115833.3$+43$5648.5 & 55 $\times$ 25 
		& 820 & 0.18 &$>$$3$$\times$$\e{20}$& 4.4 & $>$$1$$\times$$\e{4}$ \\
115833.2$+43$5648.4 & 65 $\times$ 60
		& 810 & 0.13 &$>$$3$$\times$$\e{20}$& 1.3 & $>$$9$$\times$$\e{3}$ \\
115833.1$+43$5647.9 & 110 $\times$ 70 
		& 650 & 0.21 &$>$$4$$\times$$\e{20}$& 1.9 & $>$$3$$\times$$\e{4}$ \\
115832.2$+43$5644.6 & 30 $\times$ 30 
		& 690 & 0.20 &$>$$4$$\times$$\e{20}$& 4.4 & $>$$4$$\times$$\e{3}$ \\
115832.0$+43$5644.1 & 60 $\times$  50
		& 660 & 0.19 &$>$$4$$\times$$\e{20}$& 2.3 & $>$$1$$\times$$\e{4}$ \\
115831.1$+43$5658.2 & 55 $\times$  40
		& 740 & 0.17 &$>$$3$$\times$$\e{20}$& 2.7 & $>$$1$$\times$$\e{4}$ \\	
115830.3$+43$5651.5 & 45 $\times$ 40 
		& 510 & 0.18 &$>$$4$$\times$$\e{20}$& 2.8 & $>$$4$$\times$$\e{3}$ \\
115830.8$+43$5640.2 & 130 $\times$  70
		& 500 & 0.23 &$>$$4$$\times$$\e{20}$& 2.0 & $>$$3$$\times$$\e{4}$ \\
115832.8$+43$5638.6 & 70 $\times$  60
		& 600 & 0.16 &$>$$3$$\times$$\e{20}$& 1.8 & $>$$1$$\times$$\e{4}$ \\
115830.0$+43$5648.9 & 55 $\times$  55
		& 430 & 0.25 &$>$$5$$\times$$\e{20}$& 2.9 & $>$$2$$\times$$\e{4}$ \\
\cline{1-7}\\
\multicolumn{7}{c}{NGC 4302} \\
\cline{1-7}
122141.9$+14$3506.8 & 90 $\times$  80
		& 590 & 0.14 &$>$$3$$\times$$\e{20}$& 1.1 &  $>$$3$$\times$$\e{4}$ \\
122143.1$+14$3512.4 & 110 $\times$  90
		& 900 & 0.19 &$>$$4$$\times$$\e{20}$& 1.3 & $>$$3$$\times$$\e{4}$ \\ 
122143.0$+14$3518.9 & 110 $\times$  100
		& 820 & 0.22 &$>$$4$$\times$$\e{20}$& 1.4 & $>$$6$$\times$$\e{4}$ \\
122141.9$+14$3545.4 & 90 $\times$ 65 
		& 560 & 0.33 &$>$$6$$\times$$\e{20}$& 3.2 & $>$$3$$\times$$\e{4}$ \\
122143.1$+14$3550.6 & 60 $\times$ 50 
		& 860 & 0.26 &$>$$5$$\times$$\e{20}$& 3.3 & $>$$2$$\times$$\e{4}$ \\
122141.8$+14$3601.9 & 130 $\times$  90
		& 660 & 0.21 &$>$$4$$\times$$\e{20}$& 1.5 & $>$$3$$\times$$\e{4}$ \\		
122143.1$+14$3620.4 & 90 $\times$ 70 
		& 950 & 0.17 &$>$$3$$\times$$\e{20}$& 1.5 & $>$$4$$\times$$\e{4}$ \\
122142.8$+14$3627.3 & 95 $\times$ 60 
		& 600 & 0.19 &$>$$4$$\times$$\e{20}$& 1.9 & $>$$5$$\times$$\e{4}$ \\
122141.8$+14$3633.9 &  95 $\times$  65
		& 610 & 0.31 &$>$$6$$\times$$\e{20}$& 2.0 & $>$$7$$\times$$\e{4}$ \\
122141.6$+14$3649.3 & 105 $\times$  95
		& 820 & 0.24 &$>$$5$$\times$$\e{20}$& 1.5 & $>$$4$$\times$$\e{4}$ \\
122143.2$+14$3658.4 & 95 $\times$  90
		& 1040 & 0.12 &$>$$2$$\times$$\e{20}$& 0.9 & $>$$2$$\times$$\e{4}$ \\
\enddata
\tablenotetext{a}{Cloud I.D. based on R.A. and Dec.}
\tablenotetext{b}{Projected vertical distance from the midplane.}
\tablenotetext{c}{Average apparent $V$-band extinction.}
\tablenotetext{d}{Column density derived using measured $a_{V}$ and the Milky Way
	value for $N_{\rm H}$/$A_{V}$ $\sim 1.9 \times\e{21}$ cm$^{-2}$ mag$^{-1}$.  These
	are lower limits since  $a_{V} < A_{V}$.}
\tablenotetext{e}{Mean density derived from $N_{\rm H}$ and the minor axis width.}
\tablenotetext{f}{Approximate mass based upon the estimated column
	density and projected area.  Includes a factor of 1.37 correction for He.}
\end{deluxetable}

\newpage
\begin{deluxetable}{lcccccc}[!h]
\centering
\tabletypesize{\scriptsize}
\tablenum{6} 
\tablewidth{0pc}
\tablecolumns{7}
\tablecaption{Comparison of Extraplanar Dust and DIG Properties 
	\label{table:comparison} }
\tablehead{
\colhead{Galaxy} & \colhead{$L_{FIR}$\tablenotemark{a}} & 
 \colhead{SFR\tablenotemark{b}} & \colhead{$\dot{\Sigma}_{*}$\tablenotemark{d}} &
\colhead{Dust} & \colhead{DIG} & \colhead{DIG} \\
\colhead{} & \colhead{[$10^{9}$ $L_\odot$]} & 
\colhead{[$M_\odot $yr$^{-1}$]} & \colhead{[$10^{-3} M_\odot $yr$^{-1}$ kpc$^{-2}$]} &
\colhead{Morphology} & \colhead{Morphology} & \colhead{Reference}  }
\startdata
NGC 5775 & 32.5 & 9.9 & 12.0 & Highly structured filaments\tablenotemark{e} &
	Many bright filaments & 1 \\
NGC 891  & 12.4 & 3.8 & 3.4 & Highly structured filaments & Bright filaments,
	 faint + diffuse & 2,3,4 \\
NGC 4217 & 8.1 & 2.5 & 4.6 & Highly structured filaments & Two faint patches & 5 \\
NGC 4013 & 4.8 & 1.5 & 2.8 & Highly structured filaments & Faint filaments; 
	faint + diffuse & 6 \\
NGC 4302 & 2.6\tablenotemark{c} & 0.9 & 1.4 & Large complexes + filaments & 
	Faint + diffuse & 5,6 \\
\enddata
\tablenotetext{a}{The $L_{FIR}$ values were calculated using $IRAS$ $60$ and $100$
\micron\ fluxes from Sanders \etal\ (2003).}
\tablenotetext{b}{The star formation rates were calculated following Kewley \etal\ (2002)
using the calculated $L_{FIR}$ and the Kennicutt (1998) star formation calibration. For
\ngc 5775, \ngc 891, and \ngc 4217 we adopt distances of $26.7$ Mpc, $9.6$ Mpc, and $17$
Mpc, respectively.} 
\tablenotetext{c}{The $IRAS$ images do not resolve \ngc 4302 and \ngc 4298.  We estimate
the fractional contribution to both $IRAS$ bands using the relative $100$ \micron\ fluxes
for these two galaxies from more recent Herschel images (Corbelli \etal\ 2012).}
\tablenotetext{d}{The star formation surface density, $\dot{\Sigma}_{*} \equiv$
SFR/$A_{25}$, where $A_{25} = \pi D^2_{25}$/$4$.}
\tablenotetext{e}{Description based on images from the Hubble Legacy Archive.}
\tablerefs{(1) Collins \etal\ 2000; (2) \pIII; (3) Rand, Kulkarni, \& Hester (1990); (4)
Dettmar (1990); (5) Rand (1996); (6) This work.}  
\end{deluxetable}

\clearpage

\end{document}